\documentclass[%
reprint,
preprintnumbers,
nofootinbib,
amsmath,amssymb,longbibliography,
aps
prd
]{revtex4-1}
\pdfoutput=1
\usepackage{graphicx}
\usepackage[utf8]{inputenc}
\usepackage{flushend}
\usepackage{dcolumn}
\usepackage{bm}
\usepackage{soul}
\usepackage{balance}
\usepackage[normalem]{ulem}
\usepackage[colorlinks = true,
            linkcolor = blue,
            urlcolor  = blue,
            citecolor = blue,
            anchorcolor = blue]{hyperref}
\usepackage{verbatim,subfigure}
\usepackage{color,ulem}
\usepackage{bbm}
\usepackage[english]{babel}
\usepackage{MnSymbol,wasysym}
\usepackage[utf8]{inputenc}
\usepackage{pifont}
\input Starburst.fd
\newcommand*\initfamily{\usefont{U}{Starburst}{xl}{n}}\initfamily 

\newcommand{\beq}{\begin{eqnarray}}
\newcommand{\eeq}{\end{eqnarray}}
\usepackage{amsmath}
\usepackage{tikz}
\newcommand{\cmark}{\ding{52}}%
\newcommand{\xmark}{\ding{56}}%

\usetikzlibrary{decorations.pathmorphing}
\usetikzlibrary{shapes.misc}
\tikzset{cross/.style={cross out, draw=black, minimum size=8*(#1-\pgflinewidth), inner sep=0pt, outer sep=0pt},
cross/.default={1pt}}
\usetikzlibrary{patterns,math}
\definecolor{applegreen}{rgb}{0.55, 0.71, 0.0}
\usepackage{contour}
\usepackage{xparse}
\usepackage{enumitem}
\ExplSyntaxOn
\NewDocumentCommand{\HS}{m}
 {
  \seq_set_split:Nnn \l_tmpa_seq { ~ } { #1 }
  \seq_map_inline:Nn \l_tmpa_seq { \contour{green}{##1} ~ } \unskip
 }
\ExplSyntaxOff

\definecolor{darkviolet}{rgb}{0.58, 0.0, 0.83}
\definecolor{mygreen}{rgb}{0.0, 0.5, 0.0}

\begin{document}
\preprint{\texttt{\normalsize{IFT-UAM/CSIC-25-87, APCTP Pre2025-018}}}

%
\title{Quantum Chaos Diagnostics for non-Hermitian Systems \\from Bi-Lanczos Krylov Dynamics}

\author{Matteo Baggioli,$^{1,2,3}$ Kyoung-Bum Huh,$^{2}$ Hyun-Sik Jeong,$^{4,5,6}$ Xuhao Jiang,$^{2,7,8}$ Keun-Young Kim$^{9,10}$ and Juan F. Pedraza$^{4}$ \vspace{1.6mm}}

\affiliation{\vspace{4pt}$^{1}$School of Physics and Astronomy, Shanghai Jiao Tong University, Shanghai 200240, China}
\affiliation{$^{2}$Wilczek Quantum Center, School of Physics and Astronomy, Shanghai Jiao Tong University, Shanghai 200240, China}
\affiliation{$^{3}$Shanghai Research Center for Quantum Sciences, Shanghai 201315, China}
\affiliation{$^{4}$Instituto de F\'isica Te\'orica UAM/CSIC, Calle Nicol\'as Cabrera 13-15, 28049 Madrid, Spain}
\affiliation{$^{5}$Asia Pacific Center for Theoretical Physics, Pohang 37673, Korea}
\affiliation{$^{6}$Department of Physics, Pohang University of Science and Technology, Pohang 37673, Korea}
\affiliation{$^{7}$Technical University of Munich, TUM School of Natural Sciences, Physics Department, 85748 Garching, Germany}
\affiliation{$^{8}$Faculty for Physik, Ludwig-Maximilians-Universit\"at M\"unchen, 
Schellingstraße 4, 80799, Munich, Germany}
\affiliation{$^{9}$Department of Physics and Photon Science, Gwangju Institute of Science and Technology, Gwangju 61005, Korea}
\affiliation{$^{10}$Research Center for Photon Science Technology, Gwangju Institute of Science and Technology, Gwangju 61005, Korea\vspace{1mm}}

\begin{abstract}
In Hermitian systems, Krylov complexity has emerged as a powerful diagnostic of quantum dynamics, capable of distinguishing chaotic from integrable phases, in agreement with established probes such as spectral statistics and out-of-time-order correlators. By contrast, its role in non-Hermitian settings, relevant for modeling open quantum systems, remains less understood due to the challenges posed by complex eigenvalues and the limitations of standard approaches based on orthogonality, such as singular value decomposition. Here we demonstrate that Krylov complexity, computed via the bi-Lanczos algorithm, provides a reliable probe of quantum chaos in non-Hermitian systems, clearly discriminating chaotic and integrable regimes. Our results agree with complex spectral statistics and complex spacing ratios, underscoring the robustness of the method. Universality is supported by extensive tests in the non-Hermitian Sachdev-Ye-Kitaev model and random-matrix ensembles across multiple non-Hermitian symmetry classes, with further validation provided by the non-Hermitian random-field XXZ model as a pseudo-Hermitian system.
\end{abstract}

\maketitle
%
\textbf{Introduction.}
The search for universal diagnostics of quantum chaos remains a central pursuit in the study of many-body quantum systems. Among the emerging approaches, Krylov complexity (KC) \cite{Parker:2018yvk,Balasubramanian:2022tpr,Caputa:2024vrn} has recently gained prominence as a sensitive probe of quantum dynamics. For recent comprehensive overviews, see \cite{Nandy:2024evd,Rabinovici:2025otw}.

KC quantifies how rapidly an operator or state explores the Krylov subspace generated by time evolution, thereby encoding the dynamical growth of complexity. In standard Hermitian systems, particularly in the state-based formulation, where KC is entirely determined by the energy spectrum, it has been shown to reliably distinguish chaotic from integrable phases, in agreement with traditional spectral diagnostics such as level spacing statistics and out-of-time-order correlators~\cite{Caputa:2021ori,Erdmenger:2023wjg,Scialchi:2023bmw,Camargo:2024deu,Baggioli:2024wbz,Huh:2024ytz}. 

However, realistic quantum systems are often coupled to their environment, leading to non-Hermitian effective dynamics that behave fundamentally differently from their closed counterparts~\cite{Ashida:2020dkc,Rivas_2012}. Such frameworks naturally arise in dissipative~\cite{Takasu:2020raj} or continuously monitored quantum many-body systems~\cite{Naghiloo:2019dqw}, making non-Hermitian Hamiltonians not merely a mathematical tool, but an effective description of realistic quantum dynamics where environmental coupling can reshape relaxation, chaos, and information scrambling. In such open systems, spectral analysis reveals universal signatures through complex eigenvalue statistics~\cite{Grobe:1988zz,Grobe:1989aa}: integrable systems exhibit a two-dimensional Poisson distribution, while chaotic ones follow a Ginibre distribution~\cite{Ginibre:1965zz}.
 
The universal emergence of Ginibre statistics in non-Hermitian many-body systems~\cite{Hamazaki:2018aa,Akemann:2019ab,Sa:2020ab,Li:2021aa,Garcia-Garcia:2021rle,Shivam:2023aa} supports the Grobe-Haake-Sommers (GHS) conjecture~\cite{Grobe:1988zz}, the non-Hermitian counterpart to the Berry-Tabor and Bohigas-Giannoni-Schmit conjectures~\cite{BerryTabor,Bohigas:1983er} in the context of random matrix theory (RMT)~\cite{Meh2004}. In the RMT classification, Ginibre statistics characterize class A~\cite{Hamazaki:2020kbp}; these ensembles are commonly employed as effective Hamiltonians to study non-integrability and chaos in open systems. We note, however, that the GHS does not correctly capture the onset of classical chaos in certain open quantum systems with a well-defined classical limit \cite{PhysRevLett.133.240404}. This leaves open how classical and quantum chaos should correspond~\cite{Villasenor:2025eaz}, motivating notions such as transient or steady-state chaos in open systems~\cite{Mondal:2025qdd,Rufo:2025mop} and the development of new probes~\cite{PhysRevResearch.7.013276}. Here we focus on the mutual consistency of quantum-chaos diagnostics in non-Hermitian systems, and defer questions related to the classical limit to future work. Within this framework, a complementary probe is provided by the complex spacing ratio (CSR)~\cite{Sa:2020ab}, a generalization of the $r$-parameter~\cite{Atas:2013aa} to complex spectra, which identifies chaotic behavior through cubic radial repulsion and angular anisotropy.

Given the success of state-based Krylov complexity in diagnosing quantum chaos in closed systems, a natural question arises: \textit{Can KC remain a reliable chaos indicator in open, non-Hermitian settings?} Despite growing interest in non-Hermitian dynamics, KC has not yet been systematically explored in this context. Naive extensions, such as those based on singular value decomposition~\cite{Kawabata:2023yvo,Nandy:2024mml}, face critical challenges due to the inherent complexity of non-Hermitian evolution and the biorthogonal nature of eigenstates, and thus fail to reliably discriminate between quantum chaos and integrability~\cite{Baggioli:2025ohh}.

In this work, we address this gap by adopting a biorthogonal Krylov recursion via the bi-Lanczos algorithm, which naturally captures the intrinsic non-Hermitian nature of the system. We demonstrate that KC reliably distinguishes chaotic from integrable dynamics in non-Hermitian systems, in agreement with complex spectral statistics, CSR, and Ginibre ensemble behavior. This conclusion is supported by extensive analyses of the non-Hermitian Sachdev-Ye-Kitaev (SYK) model~\cite{Garcia-Garcia:2020ttf,Su:2020quk,Liu:2020fbd,Garcia-Garcia:2021elz,Zhang:2021klq,Garcia-Garcia:2021rle,Garcia-Garcia:2022xsh,Cipolloni:2022fej}, non-Hermitian random matrix ensembles~\cite{Sa:2020ab,Hamazaki:2020kbp}, and the non-Hermitian random-field XXZ model \cite{Akemann:2024unk,Roccati:2023tug}. Together, these results establish the universality of KC as a diagnostic tool beyond the Hermitian realm and underscore its applicability to dissipative many-body systems.

%
\vspace{1.5mm}
\textbf{Bi-lanczos algorithm.}
To investigate Krylov complexity in non-Hermitian many-body systems, we use the bi-Lanczos algorithm~\cite{Gaaf_2017,TS2000,Gruning:2011aa}, an extension of the standard Lanczos procedure~\cite{Lanczos:1950zz,RecursionBook} that accommodates the absence of a orthonormal basis.

The bi-Lanczos algorithm generates two sets of orthonormal Krylov bases, $\{|\,p_n\rangle\}$ and $\{|\,q_n\rangle\}$, subject to the bi-orthonormal condition $\langle \, p_n \,|\, q_m \rangle = \delta_{nm}$. Moreover, this procedure produces three sequences of Lanczos coefficients, 
$\{a_n,b_n,c_n\}$, that encode the dynamics of the system. These coefficients appear in the three-term recurrence relations of the bi-Lanczos algorithm for a given non-Hermitian Hamiltonian $H$:
\begin{equation}\label{biL1}
\begin{split}
\!\!\!&|\,r_{n+1}\rangle = (H - a_n) \,|\,q_n\rangle - b_n \,|\,q_{n-1}  \rangle\,, \quad\,\, |\,q_{n}\rangle = c_{n}^{-1}\, |\,r_{n}\rangle \,, \\
\!\!\!&|\,l_{n+1}\rangle = (H^\dagger - a_n^*) \,|\,p_n\rangle - c_n^* \,|\,p_{n-1}  \rangle, \quad |\,p_{n}\rangle = b_{n}^*{^{-1}}\, |\,l_{n}\rangle \,,
\end{split}
\end{equation}
which iteratively generate the biorthogonal Krylov subspaces. In this basis, the non-Hermitian $H$ admits a tridiagonal representation, in which the Lanczos coefficients directly correspond to its matrix elements:
\begin{align}\label{tridiag}
T=
\begin{pmatrix}
a_0 & b_1 & 0   & \cdots & 0 \\
c_1 & a_1 & b_2 & \ddots & \vdots \\
0   & c_2 & a_2 & \ddots & 0 \\
\vdots & \ddots & \ddots & \ddots & b_{n-1} \\
0 & \cdots & 0 & c_{n-1} & a_{n-1}
\end{pmatrix}\,,
\end{align}
where the $a_n$ correspond to the main diagonal, the $b_n$ to the superdiagonal, and the $c_n$ to the subdiagonal elements. This tridiagonal form serves as the starting point for analyzing time evolution and computing Krylov complexity in non-Hermitian systems.

To compute KC, we first select two normalized initial states to seed the biorthogonal Krylov construction. In the Hermitian case, several studies have shown that the Krylov complexity computed from the thermofield double (TFD) state at infinite temperature, i.e., a maximally entangled state, exhibits a pronounced early-time peak, serving as an effective diagnostic of quantum chaos~\cite{Balasubramanian:2022tpr,Erdmenger:2023wjg,Camargo:2024deu,Baggioli:2024wbz}. Motivated by these observations, we adopt the normalized TFD state at inverse temperature $\beta$~\cite{Balasubramanian:2022tpr}, constructed using the original Hamiltonian $H$, as the initial state in our non-Hermitian setup to test whether the same characteristic peak structure, and hence the conjectured diagnostic, persists. More precisely, we take:
\begin{align}\label{}
|\,p_0\rangle = |\,q_0\rangle = |\,{\rm TFD(\beta)}\,\rangle,
\end{align}
as the initial state that seeds the bi-Lanczos algorithm, Eq.~\eqref{biL1}. The bi-orthogonal structure of the non-Hermitian Hamiltonian is defined by
\begin{align}
H |n_+\rangle = E_n |n_+\rangle\,, \qquad
H^\dagger |n_-\rangle = E_n^* |n_-\rangle\,,
\end{align}
with $\langle n_-|m_+\rangle=\delta_{nm}$~\cite{Harper:2025lav,Guo:2024wmj}. We then use the non-Hermitian TFD state
\begin{align}
|\text{TFD}(\beta)\rangle
= \frac{1}{\sqrt{Z(\beta)}}
\left(e^{-\frac{\beta H}{4}}\otimes e^{-\frac{\beta H^\dagger}{4}}\right)
\sum_n |n_+\rangle_1 |n_+\rangle_2\,,
\end{align}
where $Z(\beta)$ fixes the normalization. This state initializes the bi-Lanczos recursion, whose two-sided Gram-Schmidt procedure generates the left and right Krylov bases and the corresponding Lanczos coefficients. In the SM, we show that the resulting KC dynamics are qualitatively robust to the choice of eigenvector basis.

The precise algorithm can be summarized as follows:
\begin{itemize}
\item[(i)] Set: $b_0=c_0=0\,,\quad a_0=\langle p_0 \,|\, H\, | \,q_0 \rangle=\langle q_0 \,|\, H\, |\, p_0 \rangle\,;$
\item[(ii)] $\displaystyle
\text{For} \,\,\, n \geq 1, \text{define}:\\ 
            |\,r_{n}\rangle
                = (H - a_{n-1})\,|\,q_{n-1}\rangle - b_{n-1}\,|\,q_{n-2}\rangle, \\
            |\,l_{n}\rangle
                = (H^{\dagger} - a_{n-1}^*)\,|\,p_{n-1}\rangle - c_{n-1}^*\,|\,p_{n-2}\rangle\,;$
\item[(iii)] \text{Compute:}\,
     $c_{n} = \sqrt{|\langle r_{n}\,| \,l_{n}\rangle|},\quad b_{n}=c_{n}^{-1}\langle r_{n}\,| \,l_{n}\rangle\,;$
\item[(iv)] \text{Compute:}\, $|\,q_{n}\rangle = c_{n}^{-1}\,|\,r_{n}\rangle,\quad |p_{n}\rangle = {b^*_{n}}^{-1}\,|\,l_{n}\rangle\,;$
\item[(v)] 
\text{Replace}:\\ 
$|\,q_n\rangle \,\rightarrow\, |\,q_n\rangle 
    - \sum_{l=1}^{n-1} \langle p_l \,|\, q_l \rangle\, |\,q_l\rangle, \\[2pt]
    |\,p_n\rangle \,\rightarrow\, |\,p_n\rangle 
    - \sum_{l=1}^{n-1} \langle q_l \,|\, p_l \rangle\, |\,p_l\rangle \,;$
\item[(vi)] \text{Compute:}\, $a_{n} = \langle p_{n} \,|\, H\, |\, q_{n} \rangle\,;$
\item[(vii)] \text{If} $b_n=0$ \text{stop; otherwise go to step (ii).}
\end{itemize}
In practice, numerical instabilities may arise due to the loss of full bi-orthogonality. To prevent the accumulation of such errors, it is necessary to perform complete bi-orthogonalization twice using the Gram-Schmidt procedure (v), which ensures that the generated Krylov basis vectors remain stable at every step. For further details on the numerical validation of our bi-Lanczos algorithm, we refer the reader to the SM.

Using either of the obtained Krylov bases, ${|\,p_n\rangle}$ or ${|\,q_n\rangle}$, any time-evolved state $|\,\Psi(t)\rangle$ can be expressed as
\begin{align}\label{wavefunction}
|\,\Psi(t)\rangle
= \sum_{n=0}^{} \Phi_n^p(t) |\,p_n\rangle
=\sum_{n=0}^{} \Phi_n^q(t) |\,q_n\rangle \,,
\end{align}
with coefficients $\Phi_n^p(t)$ and $\Phi_n^q(t)$, which define Krylov complexity in the non-Hermitian setting as
\begin{align}\label{Krylovcomplexity}
C(t) \equiv \sum_n n\left| \tilde{\Phi}_n^{p*}(t) \, \tilde{\Phi}_n^q(t) \right|\,.
\end{align}
Here, the tildes denote the dynamically normalized biorthogonal bases, which ensure that the Krylov probability remains conserved~\cite{Bhattacharya:2023yec},
\begin{equation}
    P(t) = \sum_n \left| \tilde{\Phi}_n^{p*}(t) \, \tilde{\Phi}_n^q(t) \right|=1\,.
\end{equation}
In prototypical Hermitian systems, e.g., random matrices~\cite{Balasubramanian:2022tpr,Balasubramanian:2023kwd,Caputa:2024vrn,Bhattacharjee:2024yxj,Jeong:2024jjn}, quantum billiards~\cite{Hashimoto:2023swv,Huh:2023jxt,Balasubramanian:2024ghv}, and variants of the SYK model~\cite{Erdmenger:2023wjg,Chapman:2024pdw,Baggioli:2024wbz,Huh:2024ytz}, KC typically follows a characteristic dynamical pattern: an initial growth phase, a pronounced peak, and eventual saturation, with the peak often serving as a hallmark of quantum chaos.
\begin{figure*}
\centering
\begin{minipage}{0.45\textwidth}
        \centering
        \includegraphics[width=\textwidth]{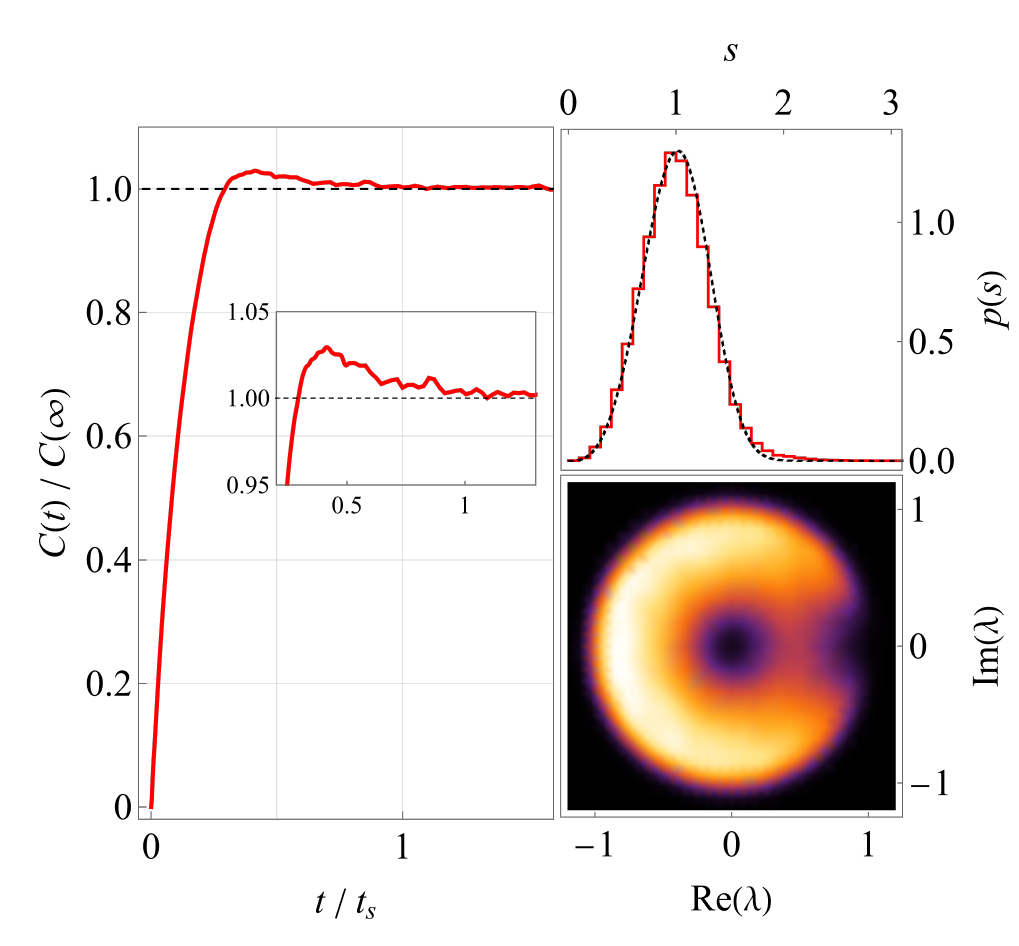}
\end{minipage}
\quad
\begin{minipage}{0.45\textwidth}
        \centering
        \includegraphics[width=\textwidth]{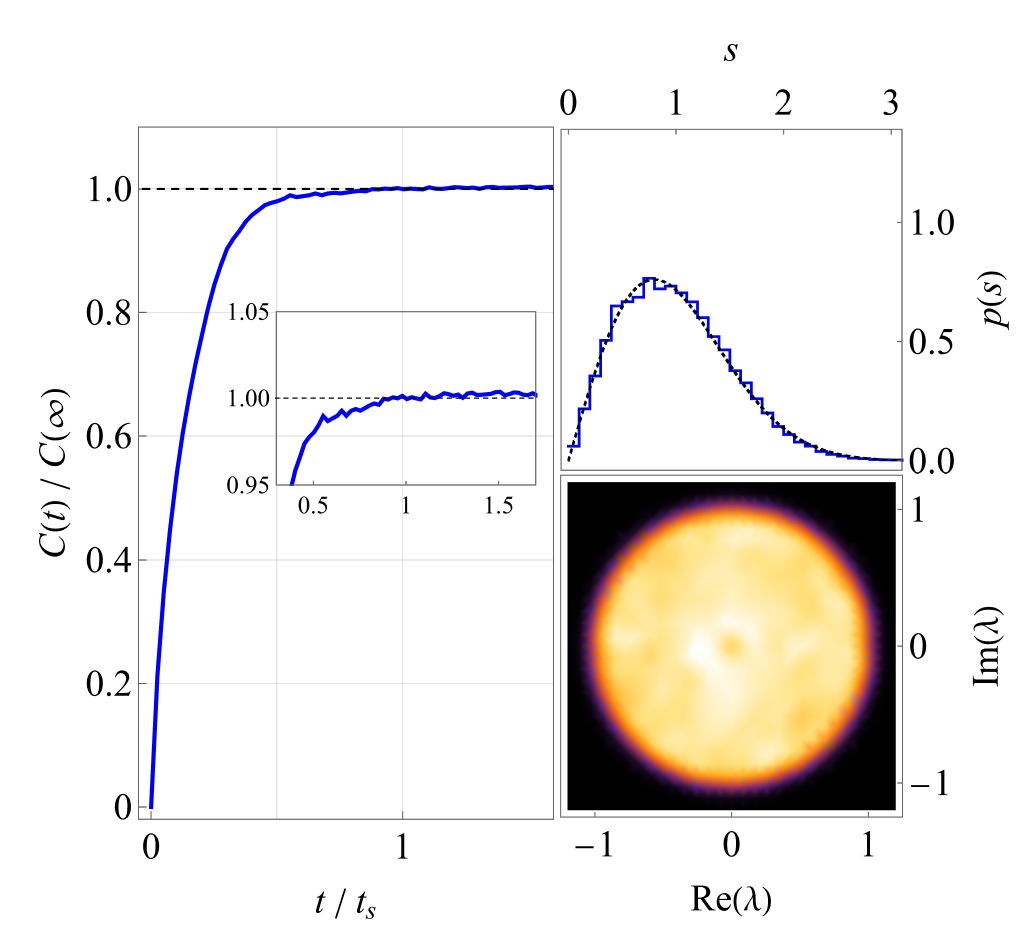}
\end{minipage}
\vspace{-0.3cm}
\caption{Unifying probes to diagnose quantum chaos in the non-Hermitian SYK model: Krylov complexity $C(t)$, complex level spacing distribution $p(s)$, and CSR in the complex plane. \textbf{Left:} For $q = 4$, clear signatures of quantum chaos are observed, including a pronounced peak in the Krylov complexity, agreement with GinUE statistics in the complex level spacing distribution, and anisotropic CSR. \textbf{Right:} In contrast, the $q = 2$ case shows no peak in Krylov complexity, a complex level spacing distribution consistent with two-dimensional Poisson statistics, and isotropic CSR.}\label{NHSYK}
\end{figure*}
In contrast, KC in non-Hermitian systems has been far less explored, leaving open the question of whether the characteristic peak persists as a universal chaos diagnostic. Progress is hindered by technical challenges: the bi-Lanczos algorithm requires the simultaneous construction of left and right Krylov bases, which are not mutually orthogonal. This doubles memory demands and amplifies numerical instabilities, often necessitating costly re-biorthogonalization. To mitigate these issues, some works restrict to complex-symmetric Hamiltonians and apply the standard Lanczos algorithm~\cite{Bhattacharya:2023yec,Medina-Guerra:2025rwa,Medina-Guerra:2025wxg}, while others retain the full bi-Lanczos approach~\cite{Chakrabarti:2025hsb} but limit themselves to smaller Hilbert spaces. Moreover, these studies have largely focused on phenomena such as the quantum Zeno effect, non-Hermitian phase transitions, or quantum Fisher information, rather than quantum chaos.

A recent exception is Ref.~\cite{Zhou:2025ozx}, which studied a disordered non-Hermitian spin chain using the bi-Lanczos algorithm and conjectured that a suppression of KC growth may signal chaotic–integrable transitions. However, no KC peak was observed in either regime, likely because the initial state was not chosen to be the TFD state. For related investigations of operator KC in open settings, see also~\cite{Liu:2022god,Bhattacharya:2022gbz,Srivatsa_2024,Bhattacharjee:2022lzy,Bhattacharya:2023zqt,Bhattacharyya:2023grv,Bhattacharjee:2023uwx}. Ref.~\cite{Bhattacharya:2023zqt} was an early step in this direction, introducing a bi-Lanczos framework for operator KC, albeit without a full numerical implementation. This operator framework is, however, conceptually different from the present work: state KC is governed by a non-Hermitian Hamiltonian, whereas operator KC is governed by a Lindbladian superoperator derived from an underlying Hermitian Hamiltonian.

%
\vspace{1.5mm}
\textbf{Non-Hermitian SYK model.}
Our first application is the non-Hermitian SYK (nHSYK) model~\cite{Garcia-Garcia:2021rle,Garcia-Garcia:2022xsh}, describing $N$ Majorana fermions in $(0+1)$ dimensions with random $q$-body interactions. It extends the original SYK model~\cite{PhysRevLett.70.3339,PhysRevX.5.041025,PhysRevD.94.106002} by including an additional non-Hermitian term. Its Hamiltonian is given by
\begin{align}\label{SYKMODEL}
    H = \sum_{i_1<i_2< \cdots < i_q }^{N}\, (J_{i_1\,i_2\,\cdots\,i_q}+i M_{i_1\,i_2\,\cdots\,i_q})\, \chi_{i_1}\, \chi_{i_2}\, \cdots \chi_{i_q} \,,
\end{align}
where $\chi_i$ are Majorana fermions satisfying $\{\chi_i,\chi_j\} = \delta_{ij}$. The random couplings $J$ and $M$ are drawn from Gaussian distributions with zero mean and variance $\langle J^2\rangle = \langle M^2\rangle = (q-1)!/N^{q-1}$. A nonzero $M$ explicitly breaks the system's Hermiticity. The nHSYK model possesses involutive symmetries generated by charge-conjugation operators~\cite{Garcia-Garcia:2021rle}. For a given pair $(N,q)$, these symmetries allow the Hamiltonian to be decomposed into symmetry blocks, thereby reducing the effective dimensionality of the Hilbert space. In the main text, we focus primarily on the case $N=22$ for both $q=4$ and $q=2$. To assess the robustness of our conclusions, we present a complementary analysis covering a broader range of $N$ and $q$ values in the \textit{End Matter} and SM.

To benchmark our results, we analyze the KC alongside the complex level spacing distribution and CSR in the non-Hermitian SYK model, where CSR is defined~\cite{Sa:2020ab} by
\begin{align}
\lambda_k = \frac{E_k^{\text{NN}} - E_k}{E_k^{\text{NNN}} - E_k}\,,
\end{align}
with $E_k$ denoting a complex eigenvalue and $E_k^{\text{NN}}$, $E_k^{\text{NNN}}$ its nearest and next-to-nearest neighbors.
Observables are averaged over $1000$ independent realizations to ensure convergence and reduce numerical error. Our results are shown in Fig.~\ref{NHSYK}. For convenience, KC is rescaled so that its late-time saturation value $C(\infty)$ equals unity, with the saturation time $t_s$ defined accordingly.

The left panel of Fig.~\ref{NHSYK} displays clear signatures of quantum chaos for $q = 4$, consistently observed across all three diagnostics. The KC, computed exactly using the bi-Lanczos algorithm, exhibits a distinct peak consistent with the spectral statistics of non-Hermitian random matrix theory. The complex eigenvalue spacing distribution $p(s)$ closely follows that of complex Gaussian non-Hermitian random matrices (GinUE)~\cite{Grobe:1988zz},
\begin{align}\label{}
   p(s) = \left( \prod_{k=1}^{\infty} \frac{\Gamma\left(1+k,s^2\right)}{k!}  \right) \sum_{j=1}^{\infty} \frac{2 s^{2j+1} e^{-s^2}}{\Gamma\left(1+j,s^2\right)} \,,
\end{align}
where $s_i = \min_{j \neq i} |E_i - E_j|$ denotes the nearest-neighbor spacing between complex eigenvalues. Here, the level spacing $s_i$ is explicitly defined as the standard Euclidean distance between two eigenenergies within the complex-energy landscape~\cite{Hamazaki:2020kbp, Garcia-Garcia:2021rle}. In addition,  $\Gamma\left(1+k,s^2\right) = \int_{s^2}^{\infty} t^k e^{-t} d t$ is defined as the incomplete gamma function. Finally, the CSR displays strong suppression at small angles, with anisotropy in the angular distribution quantified by $\langle \cos \theta \rangle \approx 0.22$, close to the GinUE value~\cite{Sa:2020ab}.

In contrast, the right panel of Fig.~\ref{NHSYK} ($q = 2$) shows clear signatures of integrability: no peak in KC, a complex eigenvalue spacing distribution following the two-dimensional Poisson distribution~\cite{Grobe:1988zz}, and a nearly isotropic CSR, with $\langle \cos \theta \rangle \approx 0.003$ for $\theta=[0,2\pi]$.

%
\vspace{1.5mm}
\textbf{Non-Hermitian random matrix universality.}
In Hermitian random matrices, Dyson’s classification into three symmetry classes (A, AI, AII) yields distinct universal behaviors characterized by Gaussian Orthogonal (GOE), Gaussian Unitary (GUE), or Gaussian Symplectic (GSE) statistics. In the non-Hermitian case, Ginibre used the same symmetry labels, all of which were found to correspond to a single universal type, the GinUE~\cite{Ginibre:1965zz} (class A). More recently, it has been shown that even in the non-Hermitian setting, three distinct universal behaviors, corresponding to classes A, $\text{AI}^\dagger$, and $\text{AII}^\dagger$, can be identified by considering an alternative symmetry involving transposition and complex conjugation~\cite{Hamazaki:2020kbp}.
\begin{figure*}[ht!]
\centering
\begin{minipage}{0.45\textwidth}
        \centering
        \includegraphics[width=\textwidth]{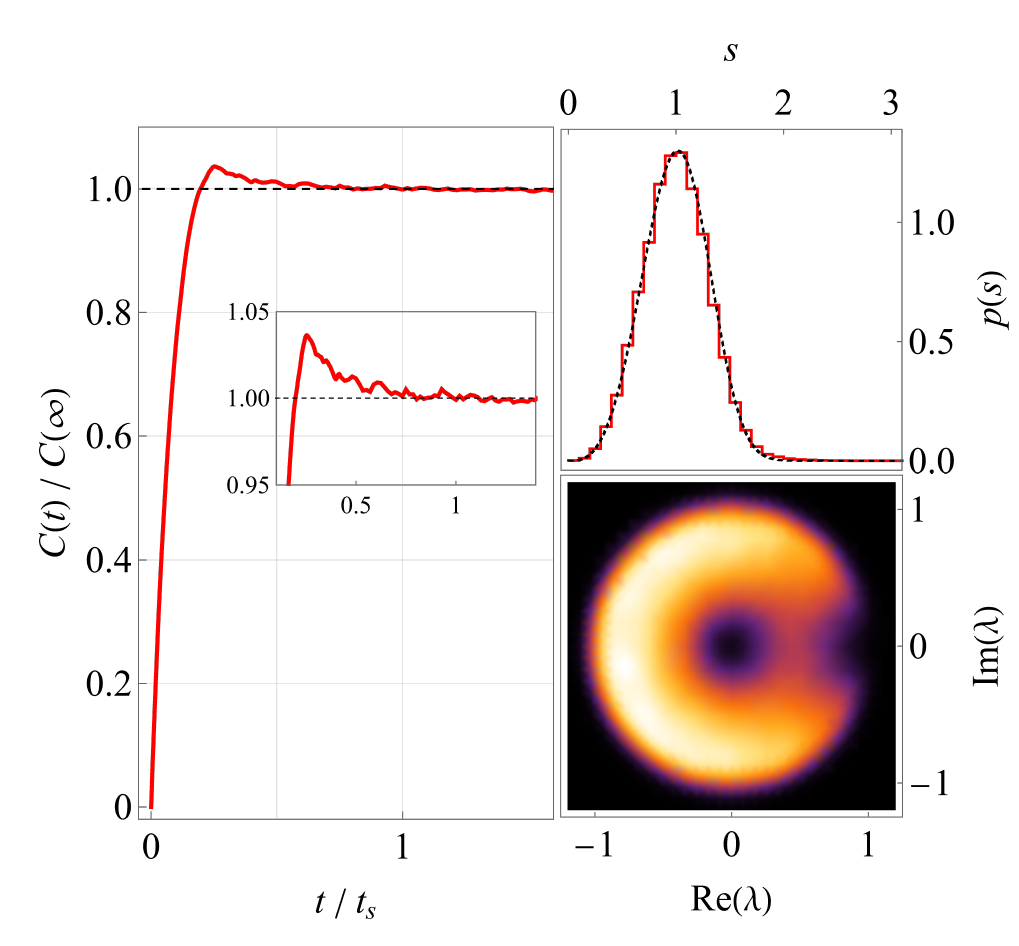}
\end{minipage}
\quad
\begin{minipage}{0.45\textwidth}
        \centering
        \includegraphics[width=\textwidth]{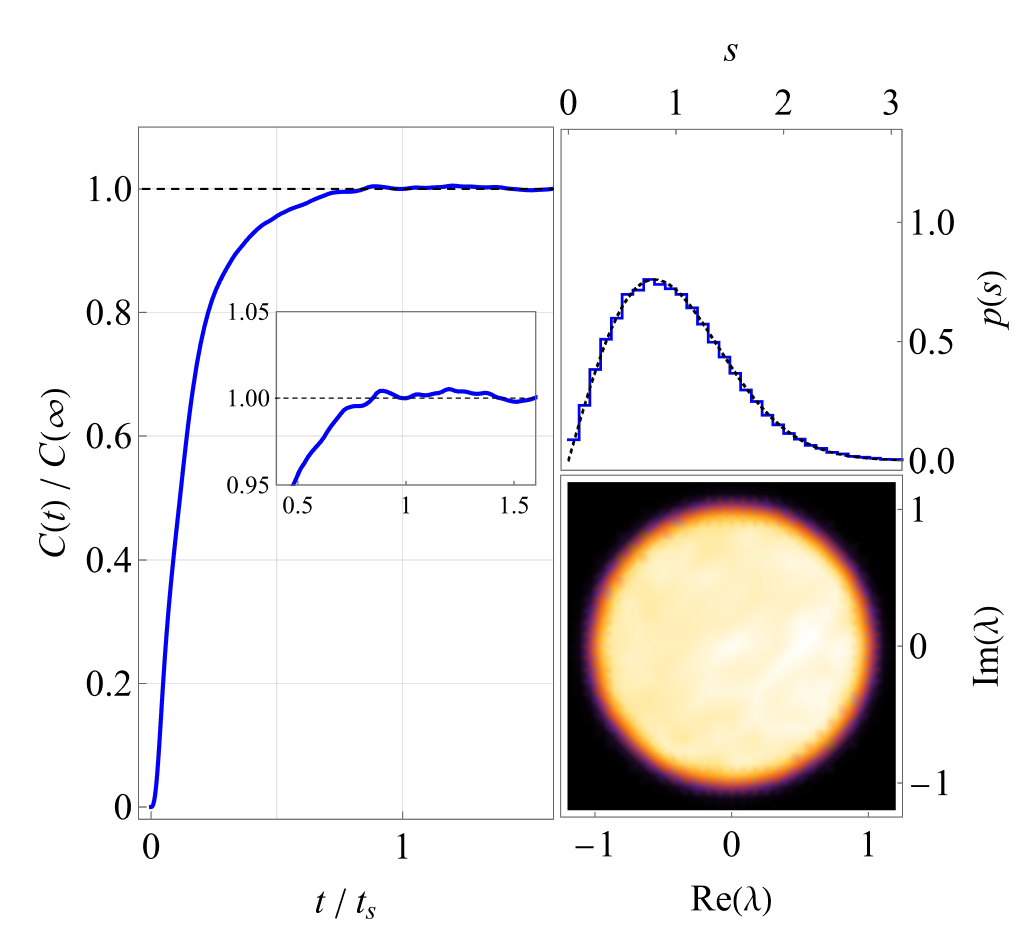}
\end{minipage}
\vspace{-0.3cm}
\caption{Unifying probes to diagnose quantum chaos in the non-Hermitian random matrix model: Krylov complexity $C(t)$, complex level spacing distribution $p(s)$, and CSR in the complex plane. \textbf{Left:} For a class~A random matrix, distinct signatures of quantum chaos are observed, including a pronounced KC peak, agreement with GinUE statistics in the complex level spacing distribution, and anisotropic CSR. \textbf{Right:} In contrast, the uncorrelated diagonal random matrix shows no KC peak, a complex level spacing distribution consistent with two-dimensional Poisson statistics, and isotropic CSR.}\label{nHRMT}
\end{figure*}
It is worth noting that all symmetry classes of non-Hermitian random matrices can be explored using the $q=4$ nHSYK model by varying the number of Majorana fermions~\cite{Garcia-Garcia:2021rle}; for example, $N=22$ yields results consistent with class A. As a second application, and to test the universality of our findings, we focus directly on non-Hermitian random matrix theories. To remain within the same symmetry class throughout this work, we choose the class A model. For comparison, and to illustrate two-dimensional Poisson statistics, we also consider a non-Hermitian random matrix with completely uncorrelated eigenvalues, which is expected to yield results consistent with the $q=2$ integrable nHSYK model.

Our findings from the non-Hermitian random matrix model under consideration closely match those from the nHSYK model, reinforcing the correspondence between them (Fig.~\ref{nHRMT}). This suggests that key features of the nHSYK model, including those revealed by the bi-Lanczos algorithm, may be universal in non-Hermitian systems with random matrix behavior. In the chaotic phase: (i) a distinct KC peak, (ii) complex spacing statistics consistent with GinUE, and (iii) anisotropic CSR are observed. In the integrable phase: (i) no KC peak, (ii) two-dimensional Poisson spacing statistics, and (iii) isotropic CSR are found. For an exhaustive investigation into the universality of the KC peak and the associated Lanczos coefficient relations, we refer the reader to the \textit{End Matter} and the SM. This includes detailed analysis of the non-Hermitian random matrix ensembles across the $\text{A}$, $\text{AI}^\dagger$, and $\text{AII}^\dagger$ symmetry classes and the
nHSYK model across multiple parameters ($q=2, 3, 4, 6$ and  $N=18, 20, 22, 24$). The results of such extensive analyses confirm the universality of our findings.

\textbf{Universality in pseudo-Hermitian models with local interactions.} 
At this stage, it is crucial to establish whether the universal signatures of quantum chaos identified via the KC peak persist in systems with local interactions. 
To this end, we performed an extensive analysis of the non-Hermitian random-field XXZ model~\cite{Akemann:2024unk,Roccati:2023tug}. 
The results, reported in the \textit{End Matter}, confirm the universality of our conjecture and establish the KC peak as a robust diagnostic of chaotic dynamics in pseudo-Hermitian systems.

%
\vspace{1.5mm}
\textbf{Discussion.}
In this work, we have shown that Krylov complexity (KC), defined through the bi-Lanczos algorithm, serves as a robust and universal probe of quantum chaos in non-Hermitian systems. The early-time KC peak, recognized as a hallmark of chaos in Hermitian systems, persists in open settings, in agreement with complex spectral diagnostics such as Ginibre statistics (GinUE) and anisotropic complex spacing ratios (CSR).

Through detailed analyses of the non-Hermitian SYK model and random matrix ensembles, we identify model-independent indicators of chaos, revealing universal behaviors across systems within the same symmetry class. Specifically, in chaotic regimes, we find a clear KC peak, GinUE-level spacing, and anisotropic CSR; in integrable regimes, these are replaced by the absence of a KC peak, two-dimensional Poisson statistics, and isotropic CSR. These results reinforce the universality of KC as a diagnostic of quantum chaos and underscore the effectiveness of biorthogonal Krylov dynamics in capturing spectral correlations and information spreading. Furthermore, recent progress in realizing high-fidelity TFD states on trapped-ion~\cite{Zhu:2019bri} and superconducting quantum-simulation platforms~\cite{Jafferis:2022crx,Wu:2018nrn} makes our TFD-based non-Hermitian Krylov framework an experimentally realistic route to probing chaotic-to-integrable transitions in open quantum systems.

Looking ahead, it will be valuable to systematically explore the degree of non-Hermiticity and how complexity measures characterize the transition from Hermitian to non-Hermitian regimes, consistent with spectral statistics. It will also be interesting to connect these ideas with dynamical probes accessible in experimental platforms that emulate non-Hermitian Hamiltonians. More broadly, investigating other Krylov-space tools, such as Krylov entropy~\cite{Barbon:2019wsy}, which quantifies the randomness of the probability distribution, and the Krylov metric~\cite{Chen:2024imd}, which tracks the geometric growth of the Krylov subspace, may uncover new signatures of chaos in open quantum systems. Together, these directions could establish bi-Lanczos Krylov dynamics as a robust, unifying framework for diagnosing quantum chaos across the full landscape of non-Hermitian many-body systems.

%
\vspace{0.15cm}
\noindent \emph{Acknowledgements.} 
We thank Debabrata Mondal and David Villaseñor for useful comments on a first version of this manuscript. MB and KBH acknowledge the support of the Foreign Young Scholars Research Fund Project (Grant No.22Z033100604). MB acknowledges the sponsorship from the Yangyang Development Fund.
HSJ and JFP are supported by the Spanish MINECO ‘Centro de Excelencia Severo Ochoa' program under grant SEV-2012-0249, the Comunidad de Madrid ‘Atracci\'on de Talento’ program (ATCAM) grant 2020-T1/TIC-20495, the Spanish Research Agency via grants CEX2020-001007-S and PID2021-123017NB-I00, funded by MCIN/AEI/10.13039/501100011033, and ERDF `A way of making Europe.'
HSJ was supported by an appointment to the JRG Program at the APCTP through the Science and Technology Promotion Fund and Lottery Fund of the Korean Government. HSJ was also supported by the Korean Local Governments -- Gyeongsangbuk-do Province and Pohang City.
KYK was supported by the Basic Science Research Program through the National Research Foundation of Korea (NRF) funded by the Ministry of Science, ICT \& Future Planning (NRF-2021R1A2C1006791), the Korea government(MSIT)(RS-2025-02311201), (RS-2024-00445164) and the framework of international cooperation program (RS-2025-02307394), the Creation of the Quantum Information Science R\& D Ecosystem (Grant No. RS-2023-NR068116). 
KYK was also supported by GIST research fund (Future leading Specialized Resarch Project, 2026, and the Regional Innovation System \& Education(RISE) program through the (Gwangju RISE Center), funded by the Ministry of Education(MOE) and the (Gwangju Metropolitan City), Republic of Korea.(2025-RISE-05-001)
HSJ would like to thank the Asia Pacific Center for Theoretical Physics (APCTP) for their hospitality during the program \textit{Holography 2025: Quantum Matter and Entanglement}, where parts of this work were undertaken.

%
\bibliographystyle{apsrev4-1}
\bibliography{Ref}

%
\section*{End Matter}\label{end}
\begin{table*}[t]
    \centering
    \renewcommand{\arraystretch}{1.3}
    \setlength{\tabcolsep}{6pt}
    \begin{tabular}{cccccc}
        \hline
                & $N = 18$ & $N = 20$ & $N = 22$ & $N = 24$ & Peak \\ 
        \hline
        \hline
        $q = 2$ & Poisson & Poisson & Poisson & Poisson & \xmark \\   
        \hline
        $q = 3$ & $\text{AII}^\dagger$ & $\text{AII}^\dagger$ & $\text{AI}^\dagger$ & $\text{AI}^\dagger$ & \cmark \\        
        $q = 4$ & $\text{A}$ & $\text{AII}^\dagger$ & $\text{A}$ & $\text{AI}^\dagger$ & \cmark \\ 
        $q = 6$ & $\text{A}$ & $\text{A}$ & $\text{A}$ & $\text{A}$ & \cmark \\ 
        \hline
    \end{tabular}
    \caption{Universality classes of the nHSYK model explored in this work. The last column summarizes whether a peak in the Krylov complexity is observed (\cmark) or not (\xmark).}
    \label{table}
\end{table*}
\textit{Appendix A: Krylov Complexity of non-Hermitian Symmetry Classes} -- The results presented in the main text focus mainly on class A. To confirm the universality of our results, here we extend our analysis of Krylov dynamics to other non-Hermitian symmetry classes. Specifically, we investigate nHSYK model and the corresponding random matrix ensembles spanning all symmetry classes A, $\text{AI}^\dagger$, and $\text{AII}^\dagger$. The symmetry classification of the nHSYK model is governed by the number of interaction $q$ and the system size $N$.

Using spectral statistics analysis (see, e.g., Ref.~\cite{Garcia-Garcia:2021rle}), we determine—as summarized in Table~\ref{table}—that each specific configuration $(N,q)$ maps onto a corresponding non-Hermitian random matrix universality class or to the Poisson ensemble. This classification enables a systematic exploration of both chaotic and integrable regimes.

As shown in Table~\ref{table}, systems belonging to all non-Hermitian symmetry classes exhibit a clear peak, whereas this feature is absent in the Poisson case. The numerical data underlying this extended analysis are provided in the Supplementary Material.

Altogether, this extended analysis verifies the universality of our conclusions across multiple symmetry classes and over a broad range of parameters and models.\\

\textit{Appendix B: Analysis of Lanczos Coefficient Ratios} -- In non-Hermitian settings, the Lanczos coefficients are found to satisfy a universal relation absent in Hermitian systems: ($1/\sqrt{2})\,|a_n|\approx|b_n|=c_n$, with the second equality holding by construction of the bi-Lanczos algorithm. In contrast, $a_n$ are weakly structured in standard closed systems~\cite{Balasubramanian:2022tpr}. 
\begin{figure}[h]
 \centering
 \vspace{1.2mm}
     {\includegraphics[width=4.1cm]{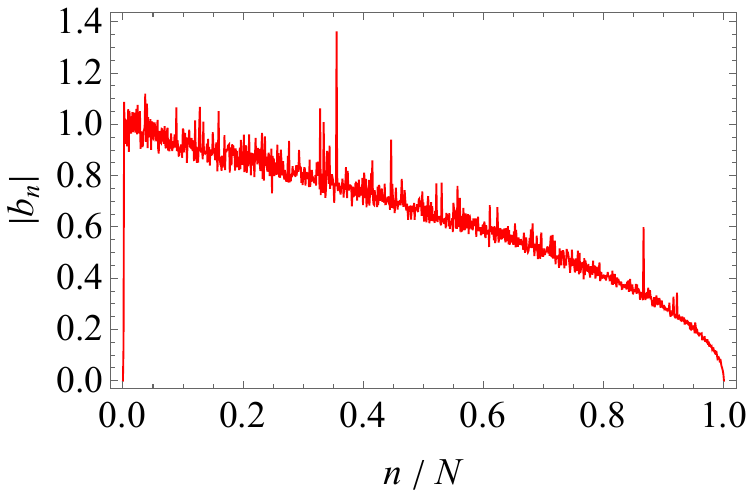}}
     {\includegraphics[width=4.1cm]{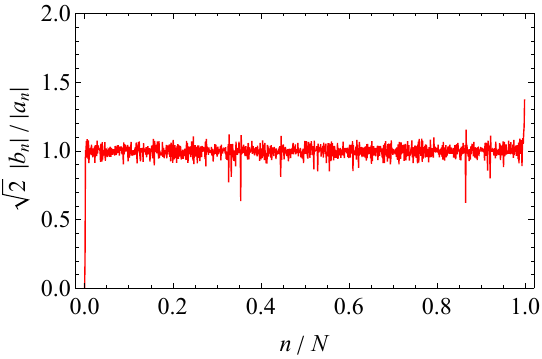}}
 \vspace{1.2mm}
     {\includegraphics[width=4.1cm]{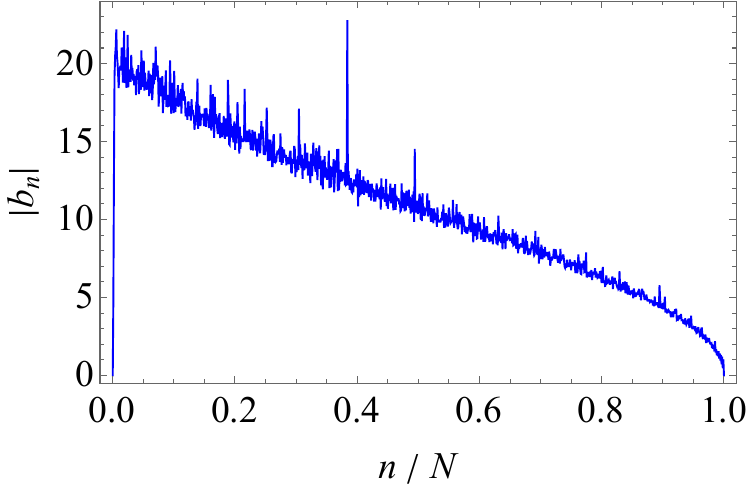}}
     {\includegraphics[width=4.1cm]{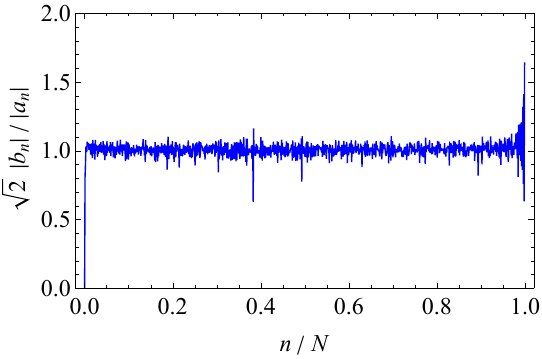}}
\caption{Lanczos coefficients for the nHSYK model. \textbf{Top:} chaotic case ($q=4$); \textbf{Bottom:} integrable case ($q=2$). Unlike the Hermitian setting, the magnitudes of the complex Lanczos coefficients are found to satisfy ($1/\sqrt{2})\,|a_n|\approx|b_n|=c_n$.} \label{Fig.Lanczos}
\end{figure}
\begin{figure}[h]
    \centering
    {\includegraphics[width=0.32\linewidth]{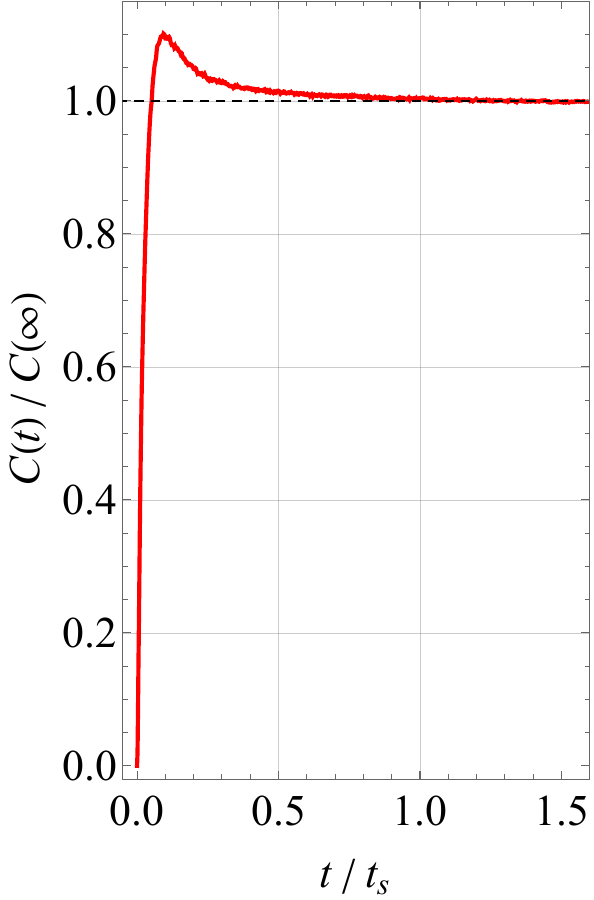}}
    {\includegraphics[width=0.32\linewidth]{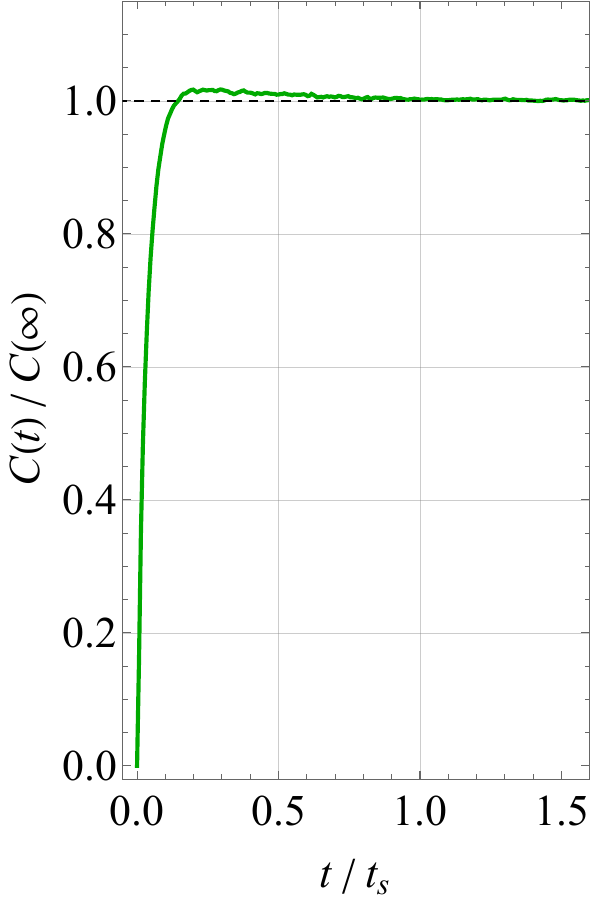}}
    {\includegraphics[width=0.32\linewidth]{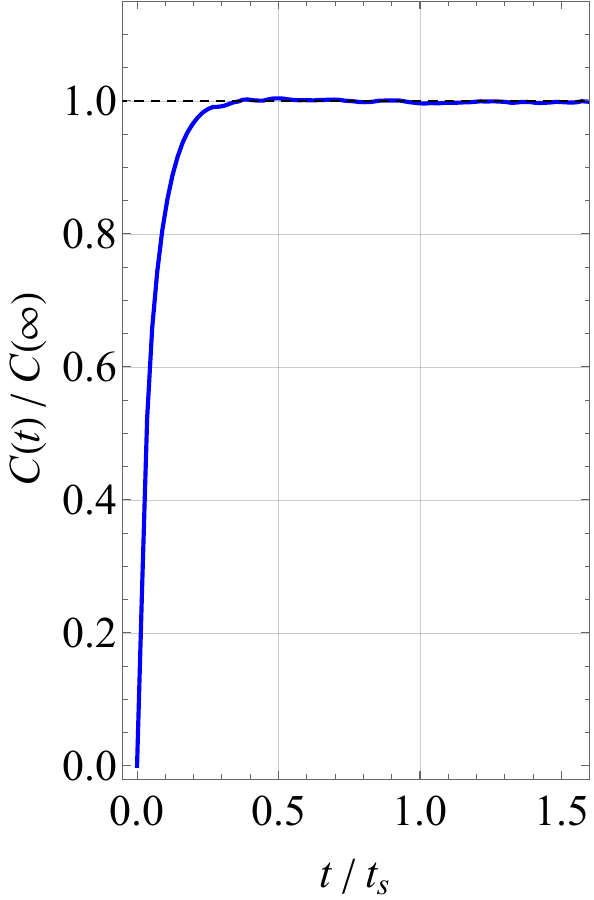}}
\caption{Krylov complexity in the random-field XXZ spin chain ($L=12$) for varying disorder strengths: $\gamma=2$ (red, chaotic), $\gamma=10$ (green, mixed phase), and $\gamma=20$ (blue, integrable). The clear emergence and suppression of the peak directly capture the transition from quantum chaotic to integrable.} \label{Fig.RXXZ}
\end{figure}
Fig.~\ref{Fig.Lanczos} shows the Lanczos coefficients of the nHSYK model with $q=4$ (red) and $q=2$ (blue), demonstrating that this relation persists in the non-Hermitian setting, regardless of whether the system is chaotic or integrable. This proportionality reflects a balanced tight-binding Krylov chain with fixed ratios between onsite and hopping amplitudes (see SM for details). In SM, we have also verified that this relation holds independently of the structure and strength of the random Hermitian and non-Hermitian couplings in our model, Eq.~\eqref{SYKMODEL}. Consistent with this structural robustness, the Lanczos-coefficient scaling $(1/\sqrt{2})|a_n|\approx |b_n|=c_n$ also holds in the Hermitian and non-Hermitian sectors of random-matrix models, as shown in the SM.

{\it Appendix C: Pseudo-Hermiticity and locality} -- 
To establish the universality of our findings against spatial locality and model perturbations, we systematically investigate the non-Hermitian random-field XXZ spin chain with local interactions, a model recently explored in Ref.~\cite{Akemann:2024unk}. The total Hamiltonian is given by
\begin{equation}
H = H_{\rm XXZ} - \frac{i}{2} \Gamma\,,
\end{equation}
where the Hermitian part is the standard XXZ spin chain,
\begin{equation}
H_{\rm XXZ} = \sum_{a=1}^L \left( S_a^x S_{a+1}^x + S_a^y S_{a+1}^y + \Delta S_a^z S_{a+1}^z \right)\,,
\end{equation}
and the dissipation is governed by 
\begin{equation}
\Gamma = \sum_{a=1}^L \gamma_a \left( S_a^z + \frac{1}{2} \mathbbm{1} \right)\,.
\end{equation}
\begin{figure}[h]
\centering
    {\includegraphics[width=0.47\linewidth]{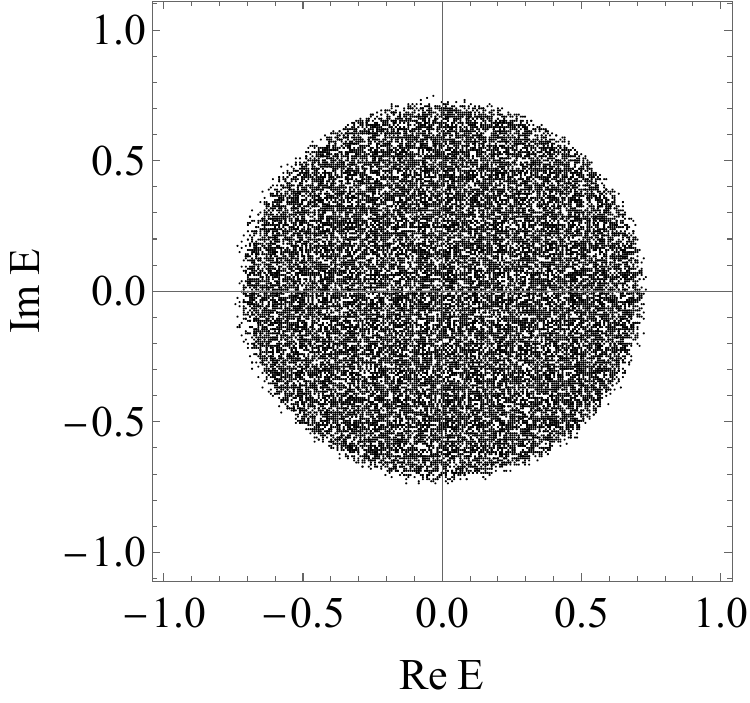}}
\quad
    {\includegraphics[width=0.47\linewidth]{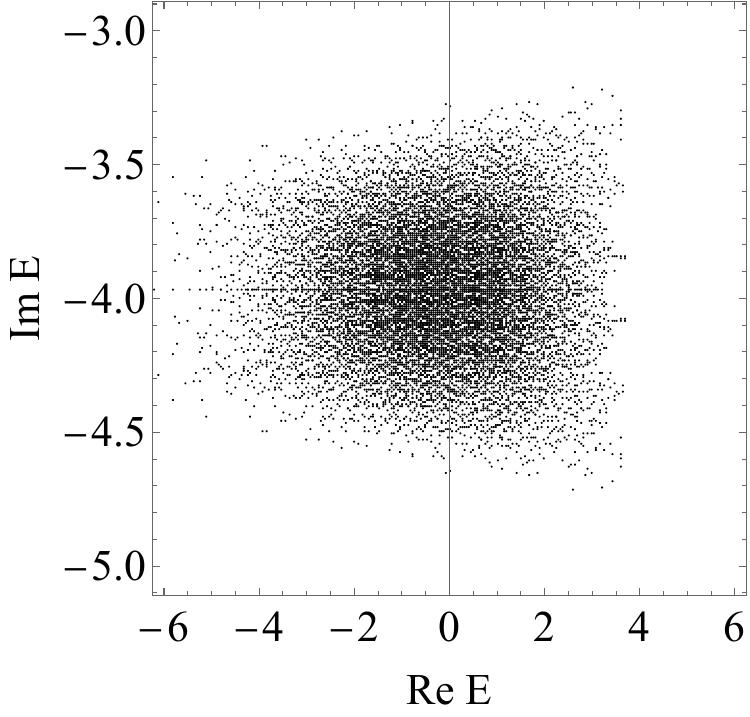}}
\caption{Complex energy level spectra. (Left) Non-Hermitian random matrix theory for class AI$^{\dagger}$. (Right) The local random-field XXZ spin chain ($L=12$, $\gamma = 2$), with its level spacing distribution adhering to class AI$^{\dagger}$.} \label{COME}
\end{figure}
Here, the disorder amplitudes $\gamma_a$ are sampled from a uniform distribution $\mathcal{U}(0, \gamma)$. This model was originally introduced in Ref.~\cite{Roccati:2023tug}, motivated by recent experimental breakthroughs demonstrating that local tunable loss terms can be successfully engineered in synthetic lattices~\cite{Lapp:2018oth}.

As established via complex spacing statistics, tuning the disorder strength $\gamma$ drives a clear transition from a chaotic phase governed by the non-Hermitian random matrix class ${\rm AI}^\dagger$ ($\gamma = 2$) to a two-dimensional Poisson integrable phase ($\gamma = 20$). 

Our numerical investigations of Krylov complexity across this transition yield two key results.
\begin{figure}[h]
\centering
    {\includegraphics[width=0.47\linewidth]{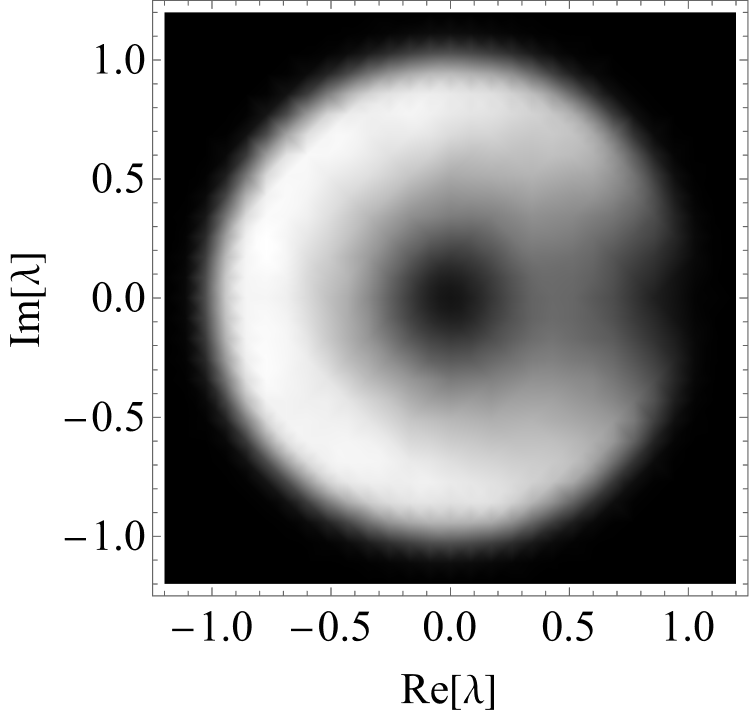}}
\quad
    {\includegraphics[width=0.47\linewidth]{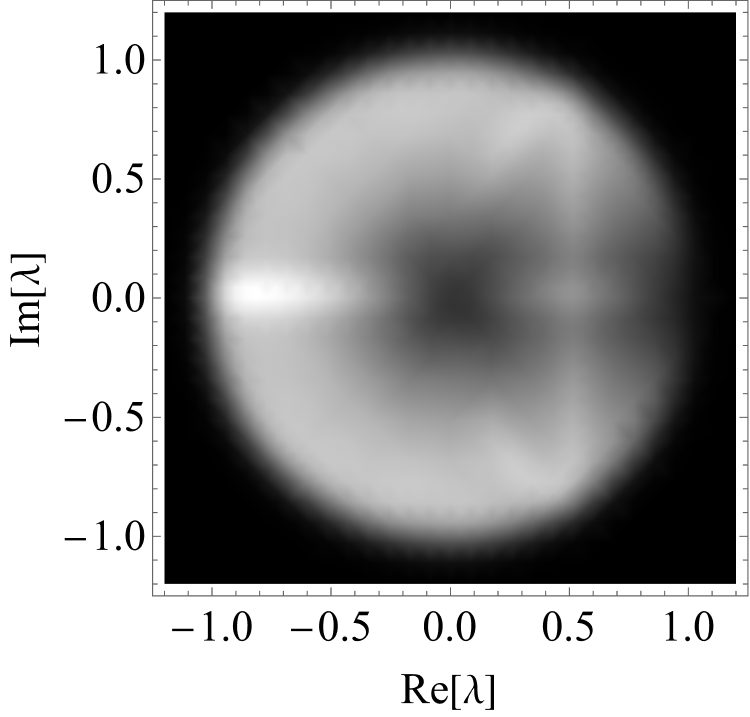}}
\caption{Complex Spacing Ratio. (Left) Non-Hermitian random matrix theory for class AI$^{\dagger}$. (Right) The local random-field XXZ spin chain ($L=12$, $\gamma = 2$), where the ``bow and arrow" feature appears.} \label{COME2}
\end{figure}
First, we confirm the robust persistence of the chaos-induced peak. As illustrated in Fig.~\ref{Fig.RXXZ}, the Krylov complexity calculations successfully capture the chaos-to-integrability transition in a highly robust manner. In the chaotic regime ($\gamma = 2$), a distinct peak structure unambiguously emerges. As the system is tuned toward the integrable regime ($\gamma = 20$), this peak is systematically suppressed and eventually disappears. This explicitly demonstrates that the peak structure is not an artifact of non-local, all-to-all interactions (such as those in the SYK model), but rather a robust indicator capable of diagnosing chaos-to-integrability transitions in local lattice models.

Second, the analysis uncovers distinct model-dependent features driven by pseudo-Hermiticity. Interestingly, the peak amplitude in the local XXZ chain is noticeably larger than that observed in pure random matrices. This feature naturally arises from the presence of an additional microscopic symmetry. Unlike generic non-Hermitian random matrices that exhibit isotropic complex spectra (see Fig.~\ref{COME}, left), the local XXZ model possesses a sharp spectral reflection symmetry parallel to the real axis due to its underlying pseudo-Hermiticity (Fig.~\ref{COME}, right)~\cite{Akemann:2024unk}. Operationally, this specific symmetry manifests as a characteristic ``bow and arrow'' geometric feature in the complex spacing ratio distribution, as verified in Fig.~\ref{COME2}.

These additional observations strongly reinforce the universality of our framework: the presence and suppression of the non-Hermitian Krylov peak serves as a robust, universal indicator of many-body quantum chaos across both local and non-local quantum systems.

Interestingly, for the local XXZ chain, the fine-grained relation $|a_n|/\sqrt{2}\approx |b_n|$, identified in our random-matrix analysis (see Appendix B in SM), no longer holds strictly. This deviation provides useful physical insight into how microscopic symmetries shape Krylov dynamics. Generic non-Hermitian random matrices have an essentially isotropic complex spectrum. By contrast, the local XXZ model exhibits a sharp spectral reflection symmetry, together with the associated ``bow-and-arrow'' structure in the spacing ratio (Fig.~\ref{COME2}), arising from its underlying pseudo-Hermiticity. This additional symmetry strongly constrains the complex matrix structure and naturally drives the diagonal ($a_n$) and off-diagonal ($b_n$) Lanczos coefficients away from the isotropic random-matrix scaling.

%
\newpage
\onecolumngrid
\appendix 
\clearpage
\renewcommand\thefigure{S\arabic{figure}}    
\setcounter{figure}{0} 
\renewcommand{\theequation}{S\arabic{equation}}
\setcounter{equation}{0}
\renewcommand{\thesubsection}{SM\arabic{subsection}}
\begin{center}
{\LARGE \textbf{Supplementary Material}}\\
\vspace{0.5cm}
{\large \textbf{``Quantum Chaos Diagnostics for non-Hermitian Systems\\ from Bi-Lanczos Krylov Dynamics"}}

\vspace{0.5cm}
{Matteo Baggioli,\, Kyoung-Bum Huh,\, Hyun-Sik Jeong,\, Xuhao Jiang,\, Keun-Young Kim\, and\, Juan F. Pedraza \vspace{1.6mm}}
\end{center}
\begin{center}
\vspace{0.0cm}
\end{center}

\vspace{0.0cm}
\onecolumngrid
This Supplementary Material presents further analyses to corroborate our findings, alongside additional details on the numerical computations with the bi-Lanczos algorithm discussed in the main text.

\tableofcontents
%
\subsection*{A. Definition of the TFD state and universality with respect to the choice of basis}
\begin{figure}[h]
        \centering
        \includegraphics[width=3.8cm]{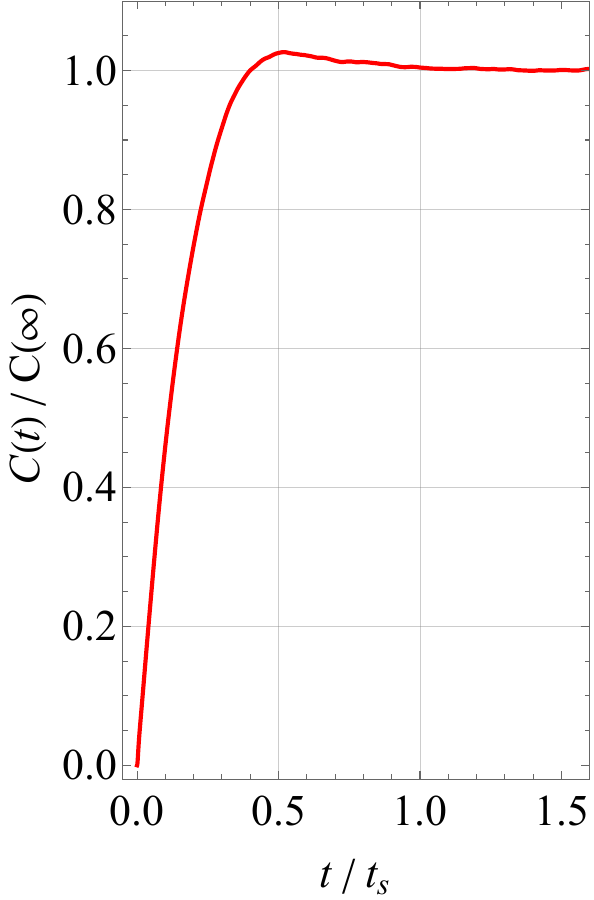}
        \centering
        \includegraphics[width=3.8cm]{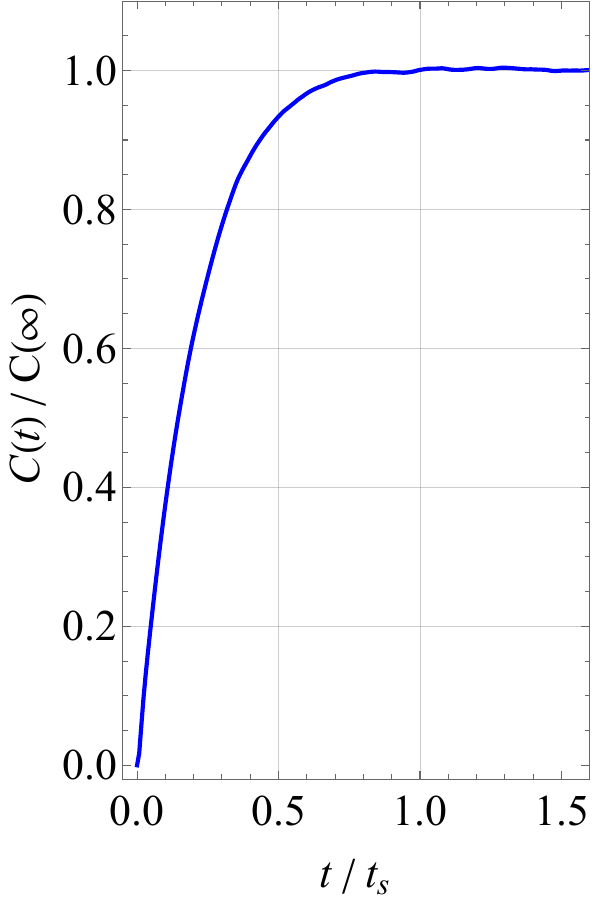}
\caption{Normalized Krylov complexity using as initial state a TFD defined using the $H^\dagger$ basis. $t_s$ refers to the saturation time and $N=22$ and $q=2,4$ (blue, red).}
\label{LTFD42}
\end{figure}

To ensure the reliability of our findings and their universal character, we have verified that the choice of basis for the TFD state does not affect our primary conclusions. As an example, the KC computed using only the $H^\dagger$ basis is displayed in
Fig.~\ref{LTFD42}. As is evident, the chaotic signature remains identifiable through the characteristic peak in the KC, which persists robustly regardless of the basis choice.
%

%
\subsection*{B. Krylov Chain Interpretation of Non-Hermitian Lanczos Coefficients}
In the Hermitian Lanczos algorithm, a single orthonormal basis $\{|\,q_n\rangle\}$ is generated by a three-term recursion. This yields a tridiagonal representation of the Hamiltonian with coefficients
\begin{align}
a_n =  \langle q_n| H |\,q_n\rangle \,, \quad b_{n+1} =  \langle q_n| H |\,q_{n+1}\rangle = c_{n+1} \,,
\end{align}
so that the resulting tridiagonal matrix form is symmetric.

On the other hand, in the non-Hermitian case, a single Krylov basis is no longer sufficient to span the full space while maintaining the tridiagonal structure, since the eigenvectors of $H$ and $H^\dagger$ form two distinct families. The bi-Lanczos procedure therefore constructs two biorthogonal bases,
\begin{align}
\{|\,q_n\rangle\} \,, \quad  \{\langle\,p_n\,|\} \,,
\end{align}
constrained by the bi-orthogonality condition $\langle \, p_n \,|\, q_m \rangle = \delta_{nm}$. This two-ladder construction preserves the tridiagonal form of the reduced Hamiltonian while maintaining stability of the recursion.

\begin{figure}[h!]
 \centering
      {\includegraphics[width=10cm]{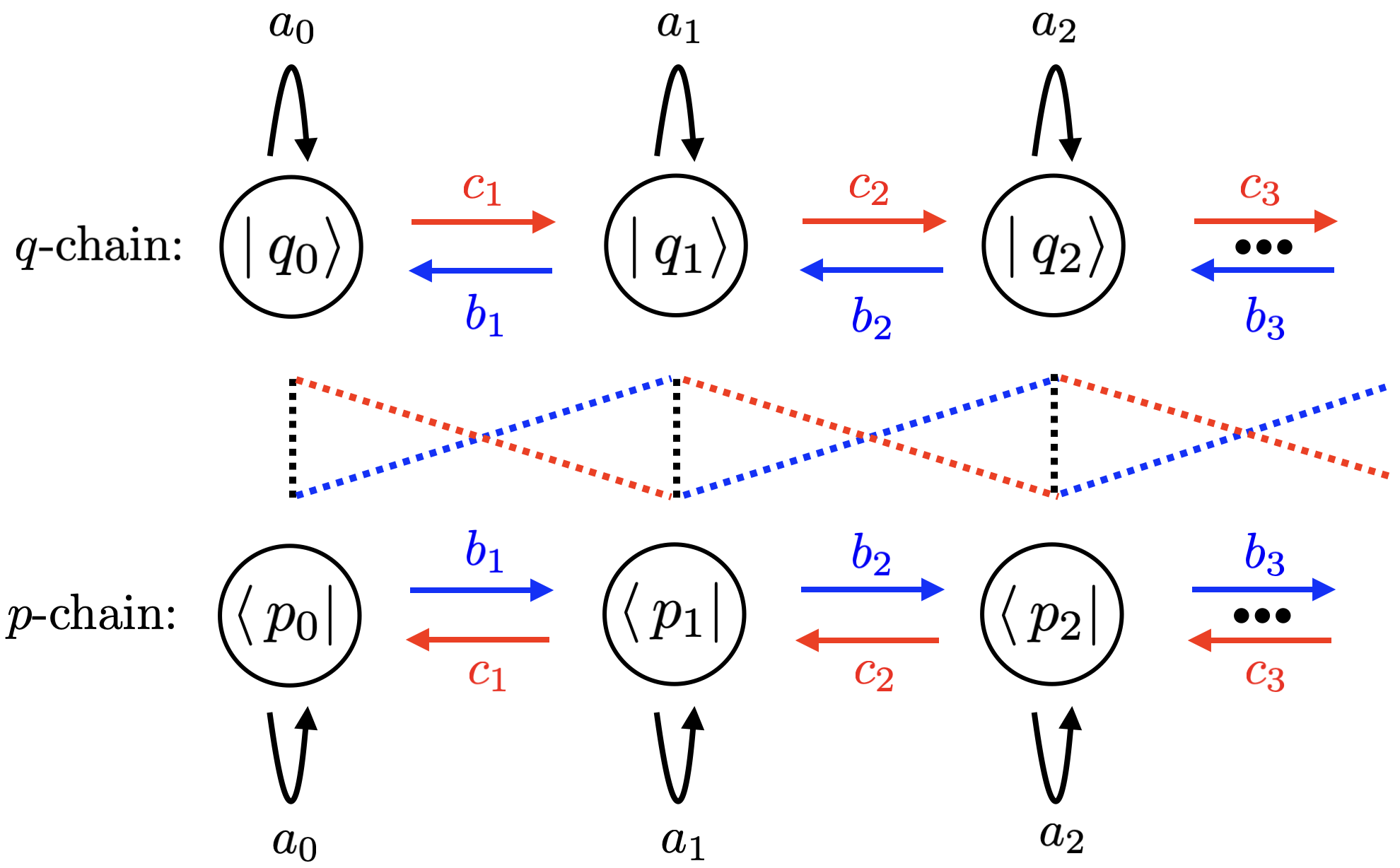}}
\caption{Schematic representation of the Hamiltonian action on the Krylov chains, Eq.~\eqref{KryChains}. Each site carries an on-site term $a_n$ (self-loop), while $b_n$ and $c_n$ are the left and right hopping amplitudes. Dashed lines indicate the overlaps $\langle p_n|H|q_m\rangle$, which determine $(a_n,b_n,c_n)$ in Eq.~\eqref{ABC}.} \label{FigChain}
\end{figure}

The associated bi-Lanczos coefficients are defined by
\begin{align}
a_n &= \langle p_n| H |\,q_{n}\rangle \,, \quad\,\,\, 
b_{n+1} = \langle p_n| H |\,q_{n+1}\rangle \,, \quad\,\,\nonumber \\
c_{n+1} &= \langle p_{n+1}| H |\,q_{n}\rangle \,, \label{ABC}
\end{align}
together with their conjugates obtained from $H^\dagger$. These coefficients govern the ``hopping" dynamics of the Krylov chains, 
where $a_n$ acts as an onsite term, while $b_n$ and $c_n$ describe nearest-neighbor transitions:
\begin{align}
&H \,|\,q_n\rangle = a_n \,|\,q_n\rangle + b_n \,|\,q_{n-1}  \rangle + c_{n+1} |\,q_{n+1}\rangle \,, \qquad
 \nonumber \\
&H^\dagger \,|\,p_n\rangle = a_n^*  \,|\,p_n\rangle + c_n^* \,|\,p_{n-1}  \rangle + b_{n+1}^* |\,p_{n+1}\rangle    \,,\label{KryChains}
\end{align}
which are obtained from the recursion relation of Eq. ({\color{blue}1}). Each chain looks like a 1D hopping chain, but we need both chains simultaneously to represent a non-Hermitian $H$. In other words, the chains are ``coupled" in the sense that all coefficients are defined as overlaps $\langle p_{n}| H |\,q_{m}\rangle$. This means the hopping strength on the $q$-chain is controlled by how the $q$-chain and $p$-chain pair up: See Fig. \ref{FigChain} for a schematic illustration of the coupled Krylov chains.\\
%

%
\subsection*{C. Empirical Observation of Lanczos coefficients}
In extensive numerical experiments with large-$N$ non-Hermitian models (including SYK-type systems and random matrix ensembles), we consistently observed the approximate relation
\begin{align}\label{APR}
    \frac{1}{\sqrt{2}}\, |a_n| \approx |b_n| = c_n \,.
\end{align}
We refer to the corresponding figures in the following section for a direct numerical confirmation of this relation across the different non-Hermitian symmetry classes.

This proportionality appears robust across both chaotic and integrable cases, pointing to a universal geometric mechanism rather than model-specific details. It suggests that many seemingly different non-Hermitian Hamiltonians reduce to the same effective recursion: a balanced tight-binding ladder with fixed ratios between onsite and hopping amplitudes.

The origin of this universality may be traced to the structure of the bi-Lanczos recursion. Unlike the Hermitian algorithm, where $a_n$ and $b_n$ emerge from independent orthogonalization steps, the bi-Lanczos scheme enforces the strong constraint of bi-orthogonality, $\langle p_n|q_m\rangle = \delta_{nm}$. This condition constrains how the action of $H$ distributes over the amplitudes among $a_n$, $b_n$, and $c_n$, forcing them to share a common ``normalization budget" to keep the bi-orthogonality condition in a balanced way.

Heuristically, one can view this as a Pythagorean-type constraint: the on-site weight $a_n$ and the hopping terms $b_n, c_n$ must balance to preserve bi-orthogonality, driving the coefficients toward the approximate relation in Eq.~\eqref{APR}. In the non-Hermitian bi-Lanczos framework, the on-site weight $a_n$ and the bi-orthogonal hopping terms ($b_n$, $c_n$) may reach a pseudo-equilibrium to preserve bi-orthogonality. Interpreting $|a_n|^2$ as the intensity of on-site persistence and $|b_n|^2 + |c_n|^2$ as the intensity of bidirectional transitions, the relation $|a_n|^2 \approx |b_n|^2 + |c_n|^2$ suggests an equipartition of dynamical weights. Given that the bi-Lanczos algorithm inherently satisfies $|b_n| = |c_n|$, this balance converges to Eq.~\eqref{APR}. While this heuristic provides physical intuition, a more rigorous analytical derivation is required to account for the complex-valued nature of the coefficients in the non-Hermitian regime.

In chaotic matrix models, the relation is expected to sharpen in the large-$N$ limit due to self-averaging, which suppresses fluctuations in the recursion coefficients. In contrast, in integrable systems the same proportionality still emerges numerically, but its robustness there appears to stem from geometric constraints of the bi-orthogonal Lanczos construction rather than statistical averaging. A systematic analytic distinction between these cases remains an open question.

%
\subsection*{D. Robustness of Lanczos coefficient relation}
\begin{figure}[]
  \centering
      {\includegraphics[width=4.1cm]{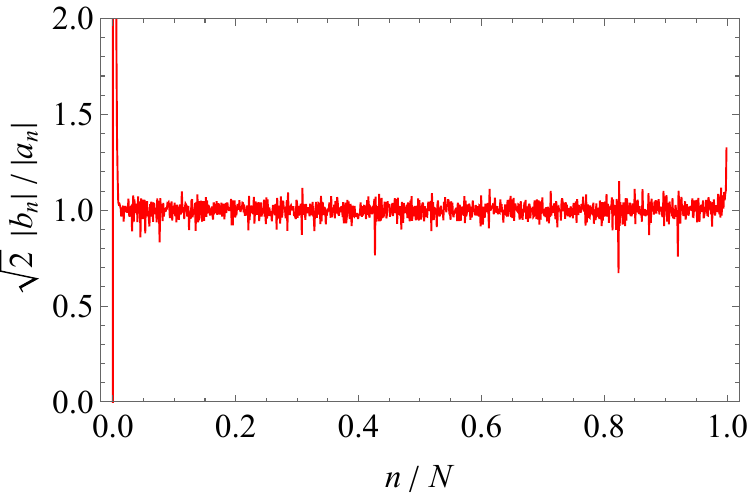}}
      {\includegraphics[width=4.1cm]{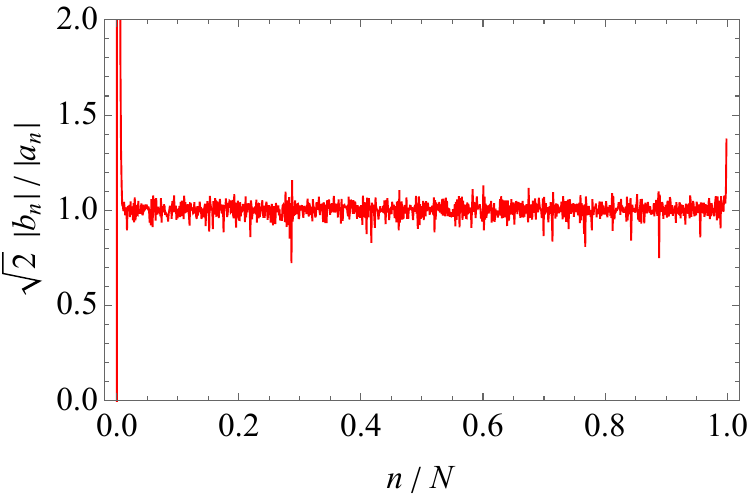}}
 \caption{Numerical verification of the relation $|a_n| \approx \sqrt{2}\,|b_n|$ under unequal variances between the Hermitian and non-Hermitian sectors. (Left) Asymmetric scaling with the non-Hermitian couplings doubled, $M_{i_1i_2\dots} = 2\,J_{i_1i_2\dots}$. (Right) Asymmetric scaling with the non-Hermitian couplings halved, $M_{i_1i_2\dots} = \frac{1}{2}\,J_{i_1i_2\dots}$.} \label{Fig.J2M}
\end{figure}
To examine whether this structure is sensitive to the relative strengths of the Hermitian and non-Hermitian sectors, Fig.~\ref{Fig.J2M} illustrates two highly asymmetric cases: in the left panel ($M = 2J$), the standard deviation of the non-Hermitian sector ($M_{i_1 i_2 \dots}$) is scaled to be twice that of the Hermitian sector ($J_{i_1 i_2 \dots}$), and in the right panel ($M = \frac{1}{2}J$), the standard deviation of the non-Hermitian sector is scaled to be half that of the Hermitian sector. As shown by the overlapping trajectories in both panels of Fig.~\ref{Fig.J2M}, the $(1/\sqrt{2})|a_n| \approx |b_n|$ scaling persists under these uneven coupling strengths, confirming that the relation reflects an intrinsic structural property governing the asymptotic behavior of the Lanczos coefficients in these non-Hermitian systems.
%
\subsection*{E. Dynamic normalization scheme}
In non-Hermitian or open quantum systems, the time evolution is inherently non-unitary. 
Consequently, the total norm of the time-evolved state vector, $\langle\Psi(t)|\Psi(t)\rangle$, is not conserved and varies continuously over time due to overall amplification or dissipation effects. 
To formulate a physically meaningful definition of state Krylov complexity, we introduce a dynamic normalization scheme. 
Operationally, at every real-time snapshot $t$, the wave function amplitudes $\Phi^{p,q}_n(t)$ across the Krylov chain are explicitly rescaled by the instantaneous total norm of the state, yielding a well-defined probability distribution:
\begin{equation}
    P_n(t) = \frac{|\Phi^{p^*}_n (t) \Phi^q_n(t)|}{\sum_m |\Phi^{p^*}_m (t) \Phi^q_m(t)|}.
\end{equation}
This continuous rescaling ensures that the total Krylov probability remains strictly conserved at unity ($\sum_n P_n(t) = 1$) for all times. 
This approach is physically justified by two crucial factors: first, it preserves the foundational meaning of complexity, $C(t) = \sum_n n P_n(t)$, which quantifies the operator or state growth as the average position of a wave packet propagating along the semi-infinite Krylov chain, an operational interpretation that would otherwise break down if the distribution $P_n(t)$ unphysically vanished or diverged due to non-unitarity; second, it systematically factors out the trivial overall growth or decay rates dictated by the imaginary parts of the spectrum, thereby allowing us to purely isolate, track, and contrast how the quantum information dynamically scrambles and spreads into higher Krylov sectors.

%
\subsection*{F. Numerical Verification of Universality across Symmetry Classes}
\begin{figure}[ht]
    \centering
    \renewcommand{\arraystretch}{1.2} 
    \setlength{\tabcolsep}{3pt}
    
    \begin{tabular}{ccccc}
        & \hspace{7mm}$\text{A}$ & \hspace{7mm}$\text{AI}^\dagger$ & \hspace{5mm}$\text{AII}^\dagger$ & \hspace{7mm}Poisson \\[2mm]
        
        \raisebox{3.0cm}{\rotatebox{90}{}}\hspace{3mm} & 
        \includegraphics[width=3.8cm]{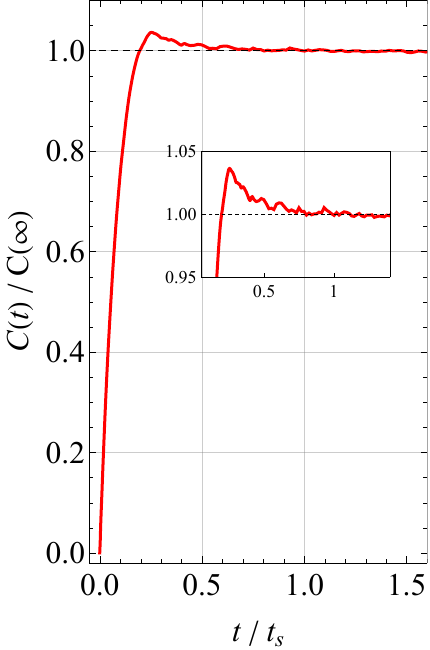} &
        \includegraphics[width=3.8cm]{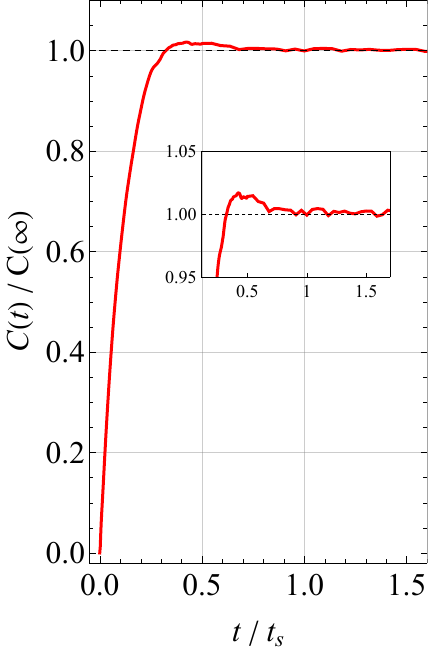} &
        \includegraphics[width=3.8cm]{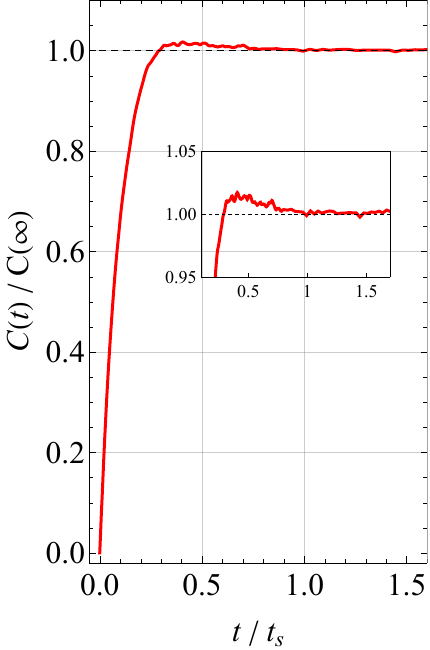} &
        \includegraphics[width=3.8cm]{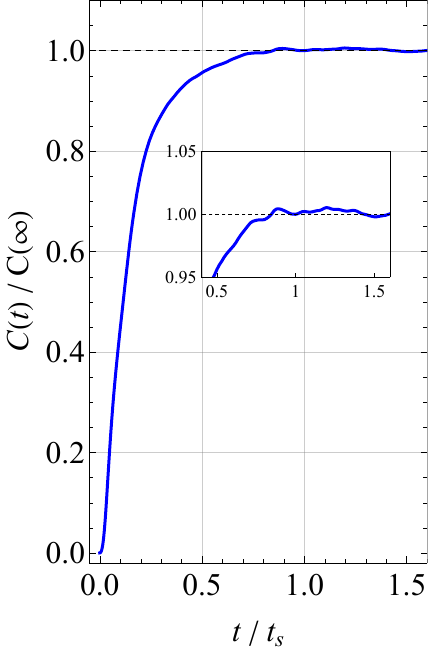} \\
    \end{tabular}
    \caption{Krylov complexity $C(t)$ of non-Hermitian random matrix model for classes $\text{A}$, $\text{AI}^\dagger$, $\text{AII}^\dagger$, and Poisson.}
    \label{KCRMT}
\end{figure}

\begin{figure}[ht]
    \centering
    \renewcommand{\arraystretch}{1.2} 
    \setlength{\tabcolsep}{3pt}
    
    \begin{tabular}{ccccc}
        & \hspace{6mm}$N=18$ & \hspace{6mm}$N=20$ & \hspace{6mm}$N=22$ & \hspace{6mm}$N=24$ \\[2mm]
        
        \raisebox{3.0cm}{\rotatebox{90}{$q=2$}}\hspace{3mm} & 
        \includegraphics[width=3.8cm]{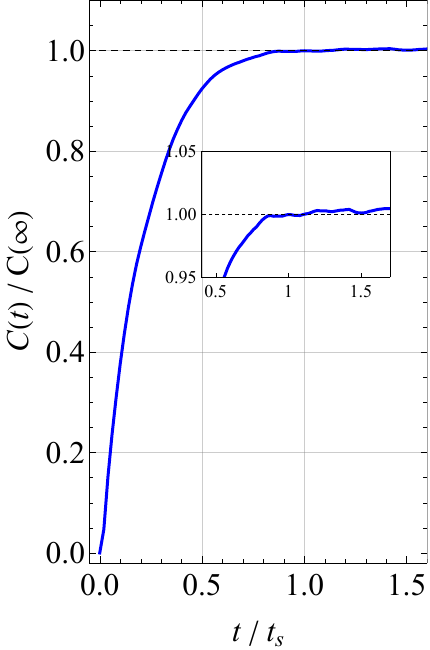} &
        \includegraphics[width=3.8cm]{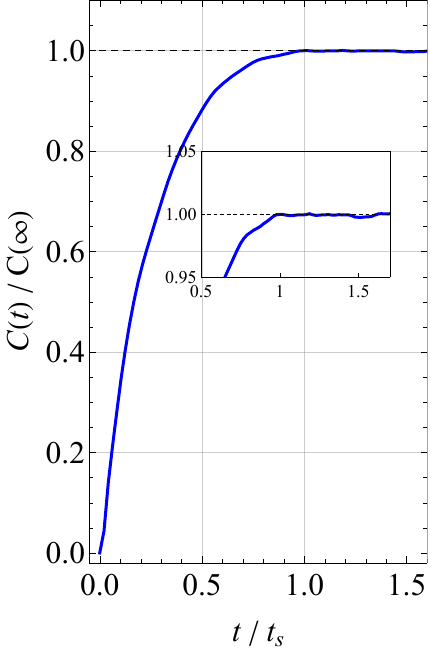} &
        \includegraphics[width=3.8cm]{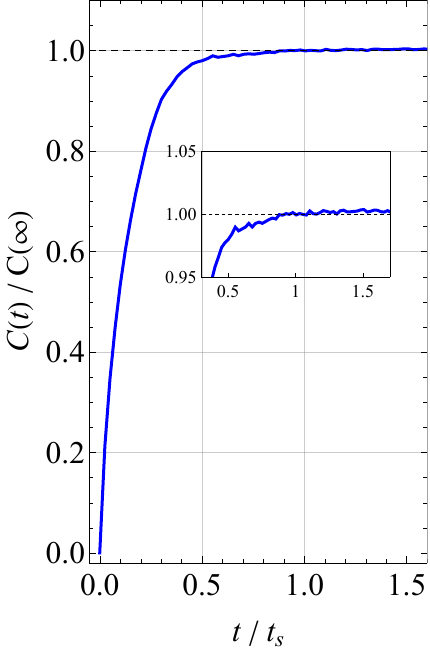} &
        \includegraphics[width=3.8cm]{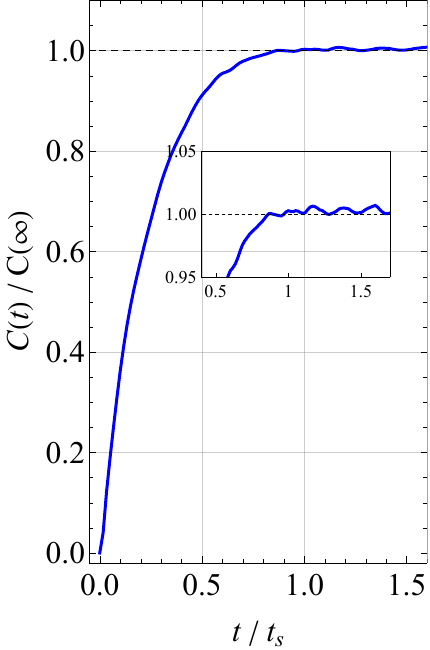} \\
    \end{tabular}
    \caption{Krylov complexity $C(t)$ of nHSYK model for $N = 18, 20, 22, 24$ and $q = 2$.}
    \label{KCq2}
\end{figure}

\begin{figure}[ht]
    \centering
    \renewcommand{\arraystretch}{1.2} 
    \setlength{\tabcolsep}{3pt}
    
    \begin{tabular}{ccccc}
        & \hspace{6mm}$N=18$ & \hspace{6mm}$N=20$ & \hspace{6mm}$N=22$ & \hspace{6mm}$N=24$ \\[2mm]
        
        \raisebox{3.0cm}{\rotatebox{90}{$q=3$}}\hspace{3mm} & 
        \includegraphics[width=3.8cm]{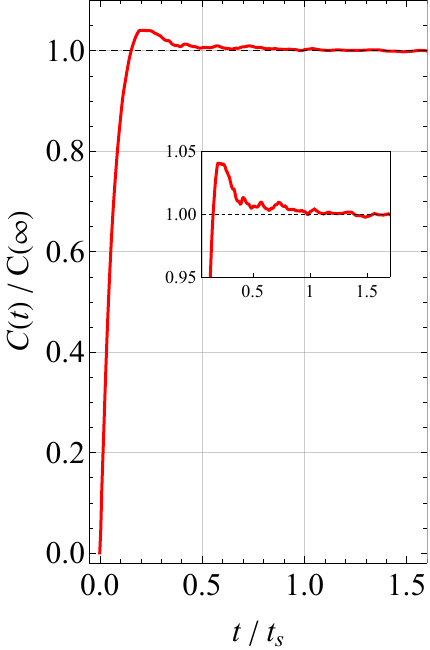} &
        \includegraphics[width=3.8cm]{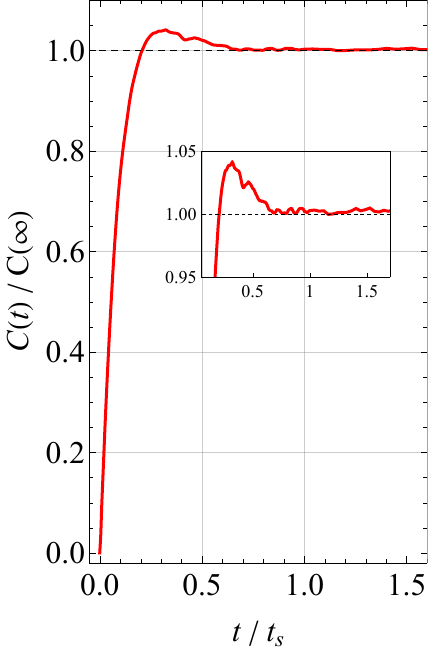} &
        \includegraphics[width=3.8cm]{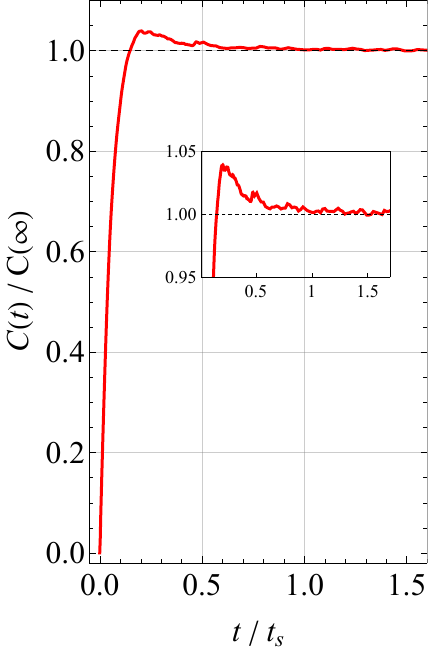} &
        \includegraphics[width=3.8cm]{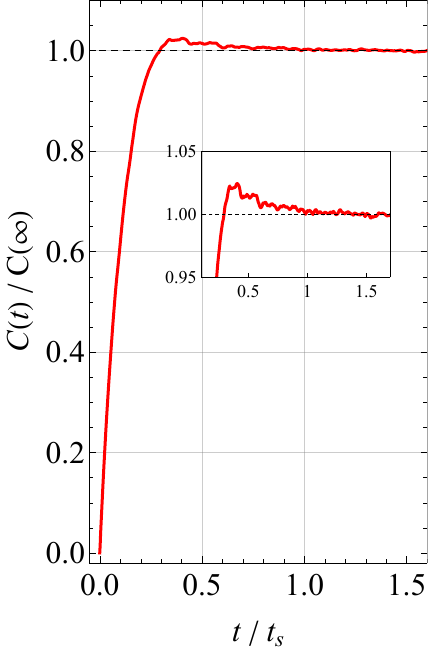} \\
        
        \raisebox{3.0cm}{\rotatebox{90}{$q=4$}}\hspace{3mm} & 
        \includegraphics[width=3.8cm]{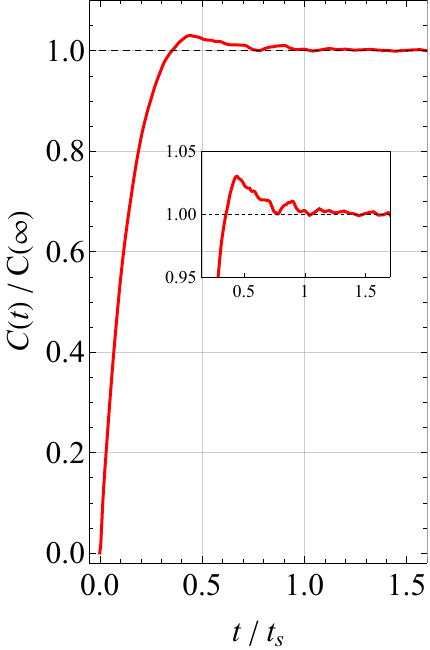} &
        \includegraphics[width=3.8cm]{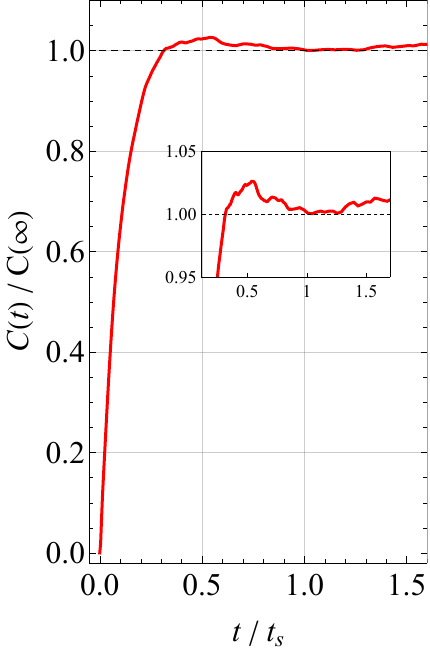} &
        \includegraphics[width=3.8cm]{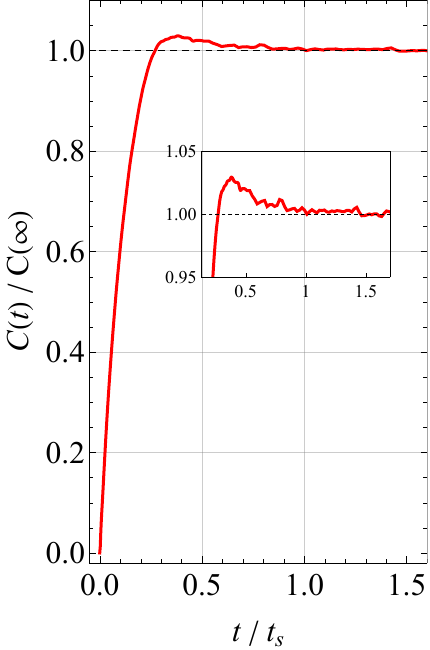} &
        \includegraphics[width=3.8cm]{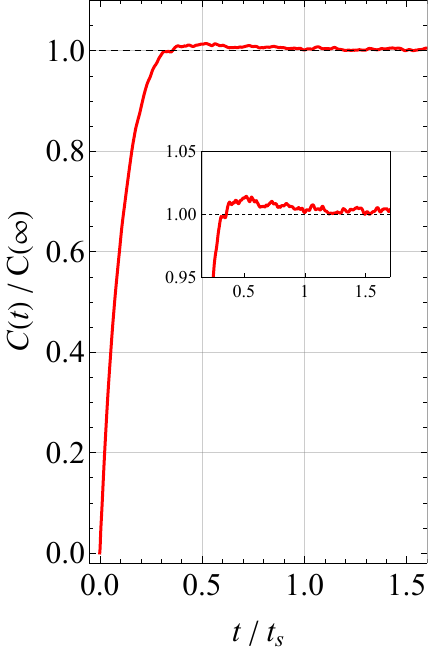} \\
        
        \raisebox{3.0cm}{\rotatebox{90}{$q=6$}}\hspace{3mm} & 
        \includegraphics[width=3.8cm]{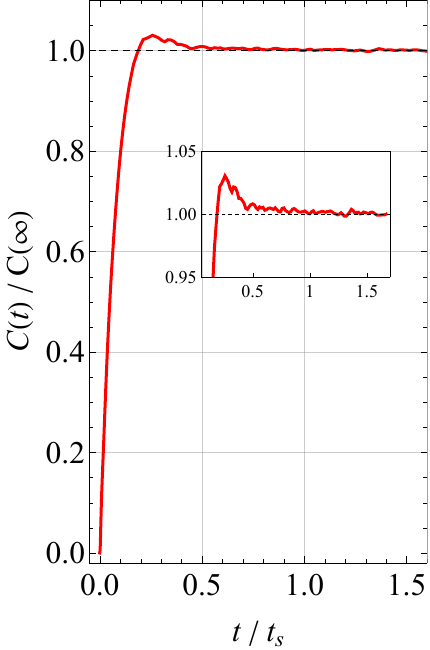} &
        \includegraphics[width=3.8cm]{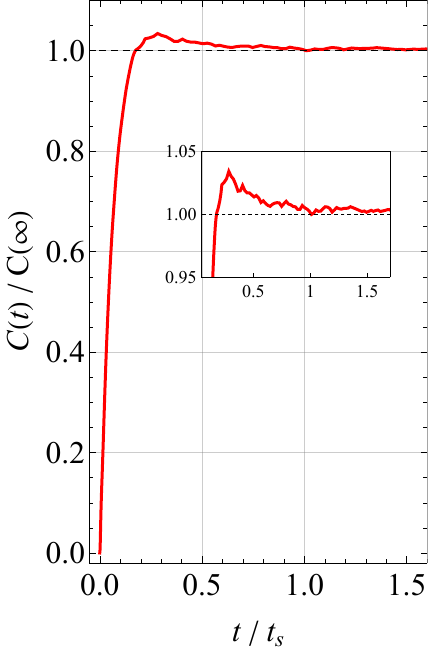} &
        \includegraphics[width=3.8cm]{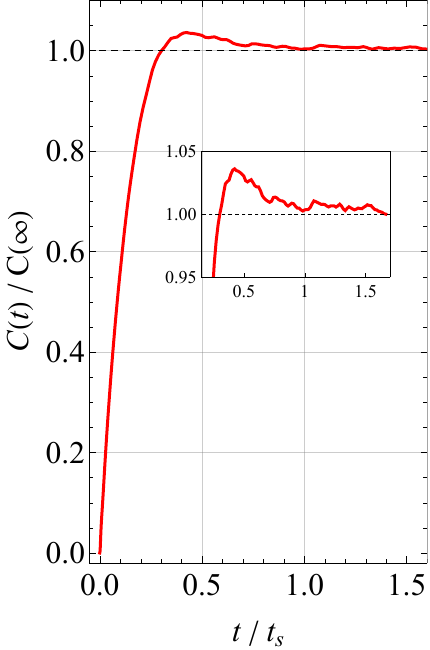} &
        \includegraphics[width=3.8cm]{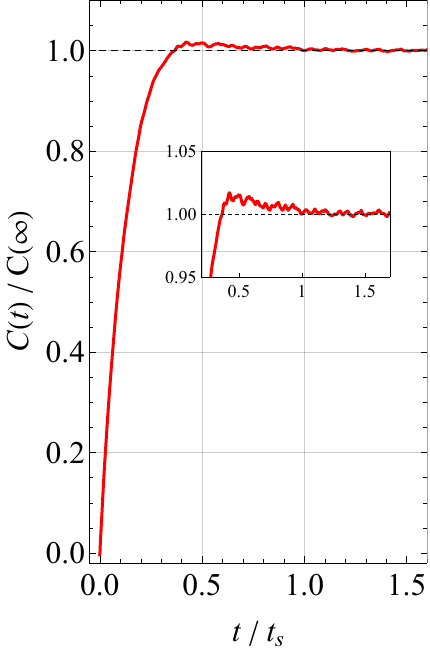} \\
    \end{tabular}
    \caption{Krylov complexity $C(t)$ of nHSYK model for $N = 18, 20, 22, 24$ and $q = 3, 4, 6$.}\label{KCq346}
\end{figure}

To complement the analysis of the various symmetry classes discussed in the main text, additional data and extended figures are presented here. 
Fig.~\ref{KCRMT} displays results for non-Hermitian random-matrix models constructed to fall into the various symmetry classes. In parallel, results for the nHSYK model corresponding to the extended analysis summarized in Table~I of the End Matter are presented in Figs.~\ref{KCq2} and \ref{KCq346}. Overall, these results confirm the universality of the statements presented in the main text.\\

\begin{figure}[ht]
    \centering
    \renewcommand{\arraystretch}{1.0} 
    \setlength{\tabcolsep}{2pt}
    
    \begin{tabular}{ccccc}
        & \hspace{5mm}$\text{A}$ & \hspace{5mm}$\text{AI}^\dagger$ & \hspace{5mm}$\text{AII}^\dagger$ & \hspace{5mm}Poisson \\[2mm]
        
        \raisebox{1cm}{\rotatebox{90}{}}\hspace{3mm} & 
        \includegraphics[width=3.9cm]{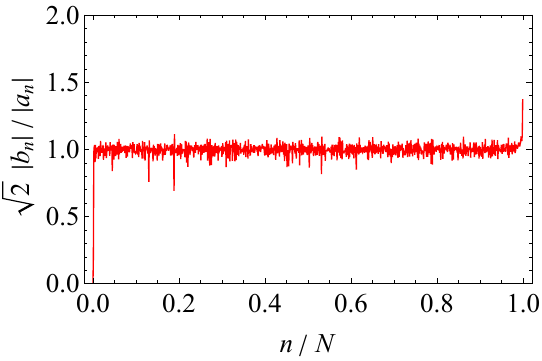} &
        \includegraphics[width=3.9cm]{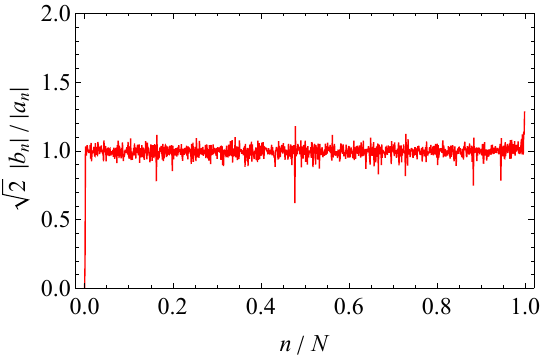} &
        \includegraphics[width=3.9cm]{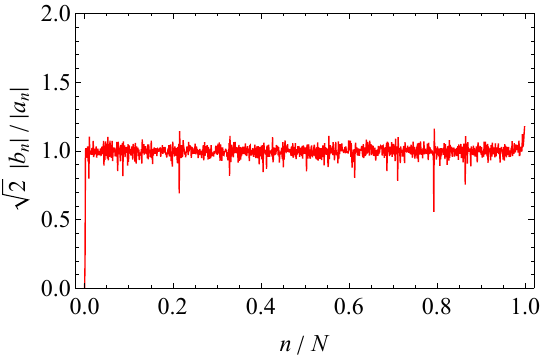} &
        \includegraphics[width=3.9cm]{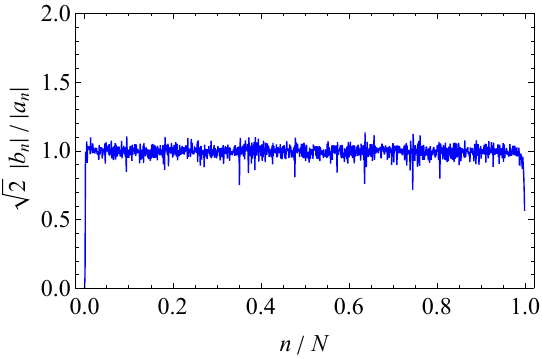} \\
    \end{tabular}
    \caption{Lanczos coefficients relation $(1/{\sqrt{2}})\, |a_n| \approx |b_n| = c_n$ of non-Hermitian random matrix model for classes $\text{A}$, $\text{AI}^\dagger$, $\text{AII}^\dagger$ and Poisson.}
    \label{LCRMT}
\end{figure}

\begin{figure}[ht]
    \centering
    \renewcommand{\arraystretch}{1.0} 
    \setlength{\tabcolsep}{2pt}
    
    \begin{tabular}{ccccc}
        & \hspace{3mm} $N=18$ & \hspace{3mm} $N=20$ & \hspace{3mm} $N=22$ & \hspace{3mm} $N=24$ \\[2mm]
        
        \raisebox{1cm}{\rotatebox{90}{$q=2$}}\hspace{3mm} & 
        \includegraphics[width=3.9cm]{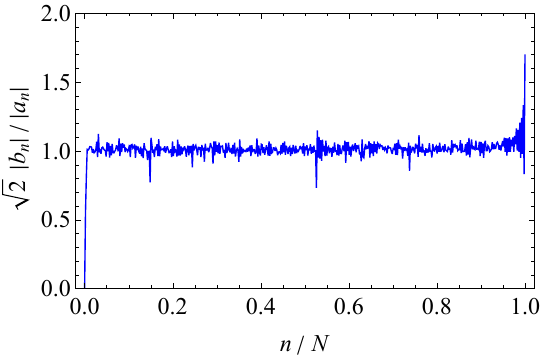} &
        \includegraphics[width=3.9cm]{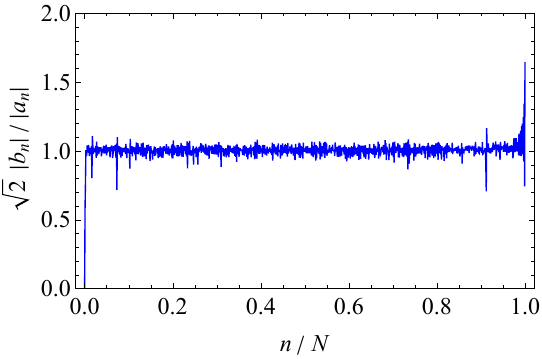} &
        \includegraphics[width=3.9cm]{nHSYKq2N22aob.pdf} &
        \includegraphics[width=3.9cm]{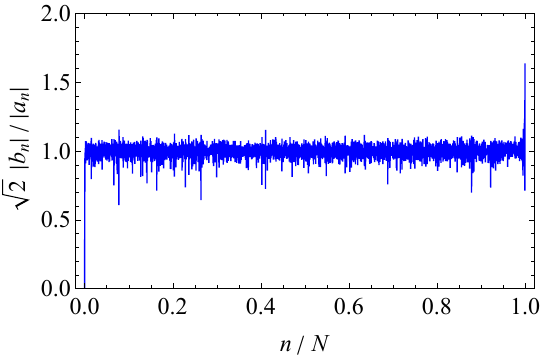} \\
    \end{tabular}
    \caption{Lanczos coefficients relation $(1/{\sqrt{2}})\, |a_n| \approx |b_n| = c_n$ of nHSYK mdoel for $N = 18, 20, 22, 24$ and $q = 2$.}
    \label{LCq2}
\end{figure}

\begin{figure}[ht]
    \centering
    \renewcommand{\arraystretch}{1.0} 
    \setlength{\tabcolsep}{2pt}
    
    \begin{tabular}{ccccc}
        & \hspace{3mm} $N=18$ & \hspace{3mm} $N=20$ & \hspace{3mm} $N=22$ & \hspace{3mm} $N=24$ \\[2mm]
        
        \raisebox{1cm}{\rotatebox{90}{$q=3$}}\hspace{3mm} & 
        \includegraphics[width=3.9cm]{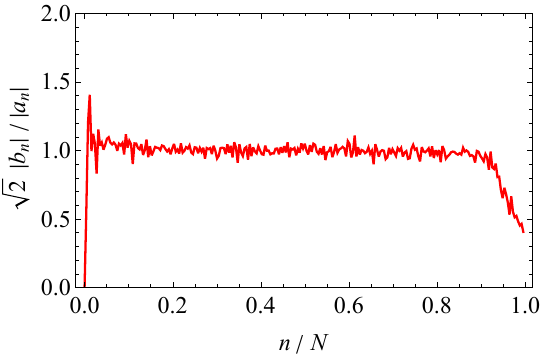} &
        \includegraphics[width=3.9cm]{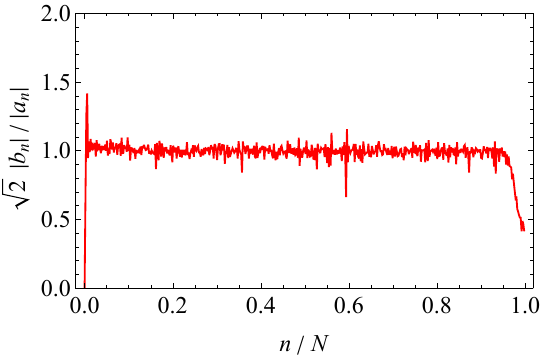} &
        \includegraphics[width=3.9cm]{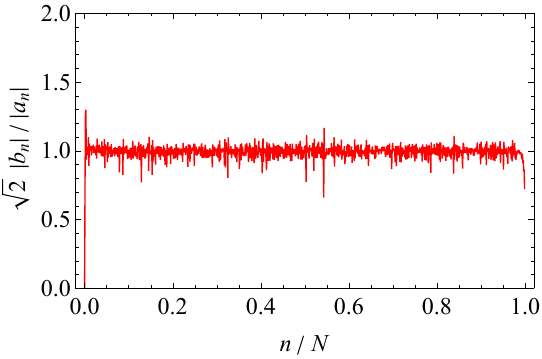} &
        \includegraphics[width=3.9cm]{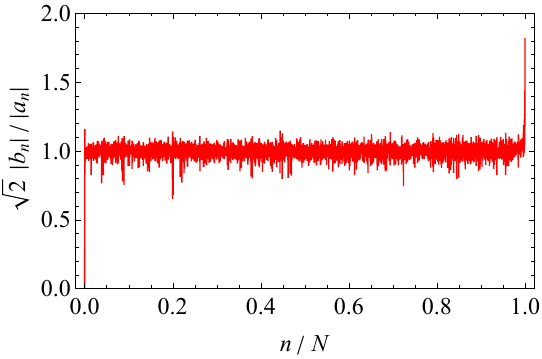} \\
        
        \raisebox{1cm}{\rotatebox{90}{$q=4$}}\hspace{3mm} & 
        \includegraphics[width=3.9cm]{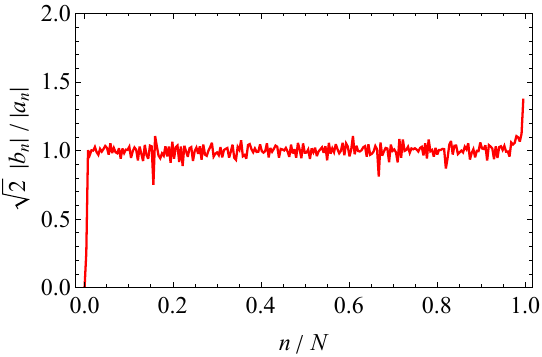} &
        \includegraphics[width=3.9cm]{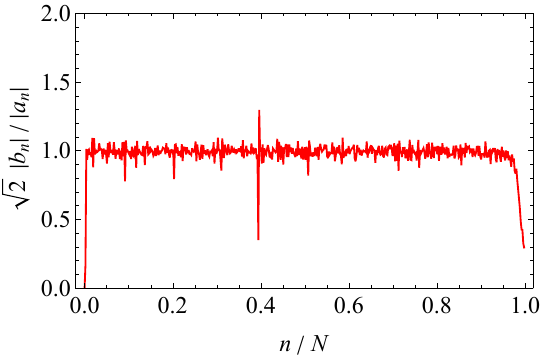} &
        \includegraphics[width=3.9cm]{nHSYKq4N22aob.pdf} &
        \includegraphics[width=3.9cm]{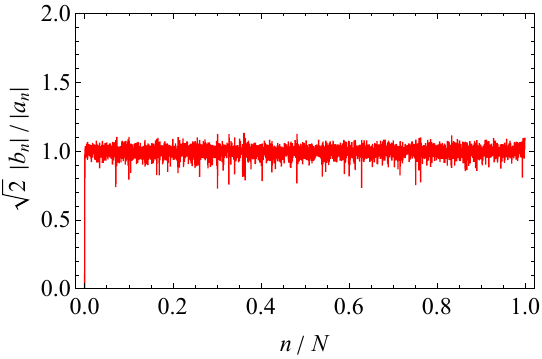} \\
        
        \raisebox{1cm}{\rotatebox{90}{$q=6$}}\hspace{3mm} & 
        \includegraphics[width=3.9cm]{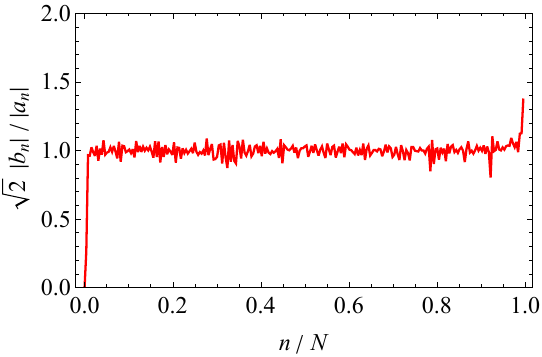} &
        \includegraphics[width=3.9cm]{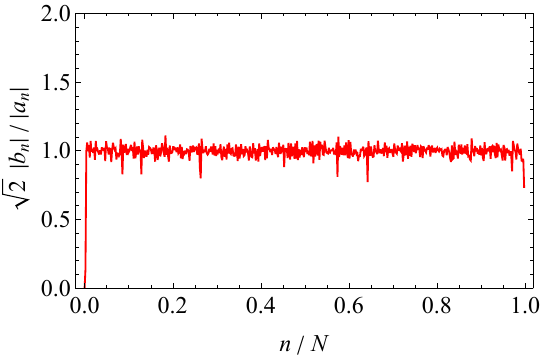} &
        \includegraphics[width=3.9cm]{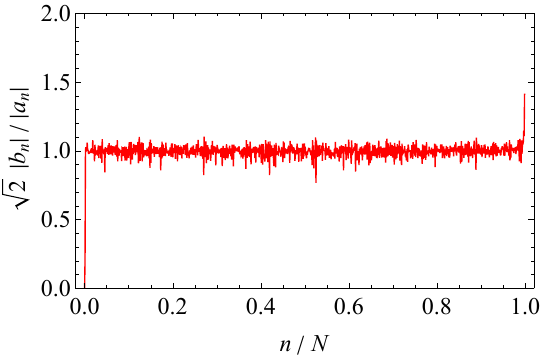} &
        \includegraphics[width=3.9cm]{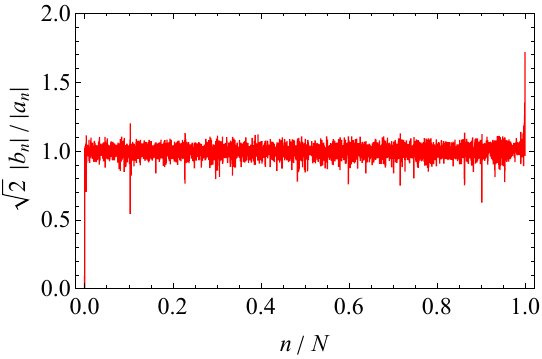} \\
    \end{tabular}
    \caption{Lanczos coefficients relation $(1/{\sqrt{2}})\, |a_n| \approx |b_n| = c_n$ of nHSYK mdoel for $N = 18, 20, 22, 24$ and $q = 3, 4, 6$.}
    \label{LCq346}
\end{figure}

In addition, we display the empirical relation among Lanczos coefficients, $(1/{\sqrt{2}}),|a_n| \approx |b_n| = c_n$, for both non-Hermitian random-matrix models and the nHSYK model in Figs.~\ref{LCRMT}, \ref{LCq2}, and \ref{LCq346}. We find that this relation remains valid across all considered symmetry classes.

\end{document}